\documentclass[5p,authoryear]{elsarticle}
\usepackage{amssymb}
\usepackage{amsmath}
\usepackage{url}
\usepackage{caption}
\usepackage{subcaption}
\usepackage{graphicx}
\usepackage{color}
\usepackage{fancyvrb}
\usepackage{microtype}
\usepackage{booktabs}
\usepackage{float}

\newfloat{listing}{ht}{lst}
\floatname{listing}{Listing}

\newfloat{xpath}{ht}{xpt}
\floatname{xpath}{XPath expression}

\makeatletter
\def\PY@reset{\let\PY@it=\relax \let\PY@bf=\relax%
    \let\PY@ul=\relax \let\PY@tc=\relax%
    \let\PY@bc=\relax \let\PY@ff=\relax}
\def\PY@tok#1{\csname PY@tok@#1\endcsname}
\def\PY@toks#1+{\ifx\relax#1\empty\else%
    \PY@tok{#1}\expandafter\PY@toks\fi}
\def\PY@do#1{\PY@bc{\PY@tc{\PY@ul{%
    \PY@it{\PY@bf{\PY@ff{#1}}}}}}}
\def\PY#1#2{\PY@reset\PY@toks#1+\relax+\PY@do{#2}}

\expandafter\def\csname PY@tok@gd\endcsname{\def\PY@tc##1{\textcolor[rgb]{0.63,0.00,0.00}{##1}}}
\expandafter\def\csname PY@tok@gu\endcsname{\let\PY@bf=\textbf\def\PY@tc##1{\textcolor[rgb]{0.50,0.00,0.50}{##1}}}
\expandafter\def\csname PY@tok@gt\endcsname{\def\PY@tc##1{\textcolor[rgb]{0.00,0.27,0.87}{##1}}}
\expandafter\def\csname PY@tok@gs\endcsname{\let\PY@bf=\textbf}
\expandafter\def\csname PY@tok@gr\endcsname{\def\PY@tc##1{\textcolor[rgb]{1.00,0.00,0.00}{##1}}}
\expandafter\def\csname PY@tok@cm\endcsname{\let\PY@it=\textit\def\PY@tc##1{\textcolor[rgb]{0.25,0.50,0.50}{##1}}}
\expandafter\def\csname PY@tok@vg\endcsname{\def\PY@tc##1{\textcolor[rgb]{0.10,0.09,0.49}{##1}}}
\expandafter\def\csname PY@tok@m\endcsname{\def\PY@tc##1{\textcolor[rgb]{0.40,0.40,0.40}{##1}}}
\expandafter\def\csname PY@tok@mh\endcsname{\def\PY@tc##1{\textcolor[rgb]{0.40,0.40,0.40}{##1}}}
\expandafter\def\csname PY@tok@go\endcsname{\def\PY@tc##1{\textcolor[rgb]{0.53,0.53,0.53}{##1}}}
\expandafter\def\csname PY@tok@ge\endcsname{\let\PY@it=\textit}
\expandafter\def\csname PY@tok@vc\endcsname{\def\PY@tc##1{\textcolor[rgb]{0.10,0.09,0.49}{##1}}}
\expandafter\def\csname PY@tok@il\endcsname{\def\PY@tc##1{\textcolor[rgb]{0.40,0.40,0.40}{##1}}}
\expandafter\def\csname PY@tok@cs\endcsname{\let\PY@it=\textit\def\PY@tc##1{\textcolor[rgb]{0.25,0.50,0.50}{##1}}}
\expandafter\def\csname PY@tok@cp\endcsname{\def\PY@tc##1{\textcolor[rgb]{0.74,0.48,0.00}{##1}}}
\expandafter\def\csname PY@tok@gi\endcsname{\def\PY@tc##1{\textcolor[rgb]{0.00,0.63,0.00}{##1}}}
\expandafter\def\csname PY@tok@gh\endcsname{\let\PY@bf=\textbf\def\PY@tc##1{\textcolor[rgb]{0.00,0.00,0.50}{##1}}}
\expandafter\def\csname PY@tok@ni\endcsname{\let\PY@bf=\textbf\def\PY@tc##1{\textcolor[rgb]{0.60,0.60,0.60}{##1}}}
\expandafter\def\csname PY@tok@nl\endcsname{\def\PY@tc##1{\textcolor[rgb]{0.63,0.63,0.00}{##1}}}
\expandafter\def\csname PY@tok@nn\endcsname{\let\PY@bf=\textbf\def\PY@tc##1{\textcolor[rgb]{0.00,0.00,1.00}{##1}}}
\expandafter\def\csname PY@tok@no\endcsname{\def\PY@tc##1{\textcolor[rgb]{0.53,0.00,0.00}{##1}}}
\expandafter\def\csname PY@tok@na\endcsname{\def\PY@tc##1{\textcolor[rgb]{0.49,0.56,0.16}{##1}}}
\expandafter\def\csname PY@tok@nb\endcsname{\def\PY@tc##1{\textcolor[rgb]{0.00,0.50,0.00}{##1}}}
\expandafter\def\csname PY@tok@nc\endcsname{\let\PY@bf=\textbf\def\PY@tc##1{\textcolor[rgb]{0.00,0.00,1.00}{##1}}}
\expandafter\def\csname PY@tok@nd\endcsname{\def\PY@tc##1{\textcolor[rgb]{0.67,0.13,1.00}{##1}}}
\expandafter\def\csname PY@tok@ne\endcsname{\let\PY@bf=\textbf\def\PY@tc##1{\textcolor[rgb]{0.82,0.25,0.23}{##1}}}
\expandafter\def\csname PY@tok@nf\endcsname{\def\PY@tc##1{\textcolor[rgb]{0.00,0.00,1.00}{##1}}}
\expandafter\def\csname PY@tok@si\endcsname{\let\PY@bf=\textbf\def\PY@tc##1{\textcolor[rgb]{0.73,0.40,0.53}{##1}}}
\expandafter\def\csname PY@tok@s2\endcsname{\def\PY@tc##1{\textcolor[rgb]{0.73,0.13,0.13}{##1}}}
\expandafter\def\csname PY@tok@vi\endcsname{\def\PY@tc##1{\textcolor[rgb]{0.10,0.09,0.49}{##1}}}
\expandafter\def\csname PY@tok@nt\endcsname{\let\PY@bf=\textbf\def\PY@tc##1{\textcolor[rgb]{0.00,0.50,0.00}{##1}}}
\expandafter\def\csname PY@tok@nv\endcsname{\def\PY@tc##1{\textcolor[rgb]{0.10,0.09,0.49}{##1}}}
\expandafter\def\csname PY@tok@s1\endcsname{\def\PY@tc##1{\textcolor[rgb]{0.73,0.13,0.13}{##1}}}
\expandafter\def\csname PY@tok@sh\endcsname{\def\PY@tc##1{\textcolor[rgb]{0.73,0.13,0.13}{##1}}}
\expandafter\def\csname PY@tok@sc\endcsname{\def\PY@tc##1{\textcolor[rgb]{0.73,0.13,0.13}{##1}}}
\expandafter\def\csname PY@tok@sx\endcsname{\def\PY@tc##1{\textcolor[rgb]{0.00,0.50,0.00}{##1}}}
\expandafter\def\csname PY@tok@bp\endcsname{\def\PY@tc##1{\textcolor[rgb]{0.00,0.50,0.00}{##1}}}
\expandafter\def\csname PY@tok@c1\endcsname{\let\PY@it=\textit\def\PY@tc##1{\textcolor[rgb]{0.25,0.50,0.50}{##1}}}
\expandafter\def\csname PY@tok@kc\endcsname{\let\PY@bf=\textbf\def\PY@tc##1{\textcolor[rgb]{0.00,0.50,0.00}{##1}}}
\expandafter\def\csname PY@tok@c\endcsname{\let\PY@it=\textit\def\PY@tc##1{\textcolor[rgb]{0.25,0.50,0.50}{##1}}}
\expandafter\def\csname PY@tok@mf\endcsname{\def\PY@tc##1{\textcolor[rgb]{0.40,0.40,0.40}{##1}}}
\expandafter\def\csname PY@tok@err\endcsname{\def\PY@bc##1{\setlength{\fboxsep}{0pt}\fcolorbox[rgb]{1.00,0.00,0.00}{1,1,1}{\strut ##1}}}
\expandafter\def\csname PY@tok@kd\endcsname{\let\PY@bf=\textbf\def\PY@tc##1{\textcolor[rgb]{0.00,0.50,0.00}{##1}}}
\expandafter\def\csname PY@tok@ss\endcsname{\def\PY@tc##1{\textcolor[rgb]{0.10,0.09,0.49}{##1}}}
\expandafter\def\csname PY@tok@sr\endcsname{\def\PY@tc##1{\textcolor[rgb]{0.73,0.40,0.53}{##1}}}
\expandafter\def\csname PY@tok@mo\endcsname{\def\PY@tc##1{\textcolor[rgb]{0.40,0.40,0.40}{##1}}}
\expandafter\def\csname PY@tok@kn\endcsname{\let\PY@bf=\textbf\def\PY@tc##1{\textcolor[rgb]{0.00,0.50,0.00}{##1}}}
\expandafter\def\csname PY@tok@mi\endcsname{\def\PY@tc##1{\textcolor[rgb]{0.40,0.40,0.40}{##1}}}
\expandafter\def\csname PY@tok@gp\endcsname{\let\PY@bf=\textbf\def\PY@tc##1{\textcolor[rgb]{0.00,0.00,0.50}{##1}}}
\expandafter\def\csname PY@tok@o\endcsname{\def\PY@tc##1{\textcolor[rgb]{0.40,0.40,0.40}{##1}}}
\expandafter\def\csname PY@tok@kr\endcsname{\let\PY@bf=\textbf\def\PY@tc##1{\textcolor[rgb]{0.00,0.50,0.00}{##1}}}
\expandafter\def\csname PY@tok@s\endcsname{\def\PY@tc##1{\textcolor[rgb]{0.73,0.13,0.13}{##1}}}
\expandafter\def\csname PY@tok@kp\endcsname{\def\PY@tc##1{\textcolor[rgb]{0.00,0.50,0.00}{##1}}}
\expandafter\def\csname PY@tok@w\endcsname{\def\PY@tc##1{\textcolor[rgb]{0.73,0.73,0.73}{##1}}}
\expandafter\def\csname PY@tok@kt\endcsname{\def\PY@tc##1{\textcolor[rgb]{0.69,0.00,0.25}{##1}}}
\expandafter\def\csname PY@tok@ow\endcsname{\let\PY@bf=\textbf\def\PY@tc##1{\textcolor[rgb]{0.67,0.13,1.00}{##1}}}
\expandafter\def\csname PY@tok@sb\endcsname{\def\PY@tc##1{\textcolor[rgb]{0.73,0.13,0.13}{##1}}}
\expandafter\def\csname PY@tok@k\endcsname{\let\PY@bf=\textbf\def\PY@tc##1{\textcolor[rgb]{0.00,0.50,0.00}{##1}}}
\expandafter\def\csname PY@tok@se\endcsname{\let\PY@bf=\textbf\def\PY@tc##1{\textcolor[rgb]{0.73,0.40,0.13}{##1}}}
\expandafter\def\csname PY@tok@sd\endcsname{\let\PY@it=\textit\def\PY@tc##1{\textcolor[rgb]{0.73,0.13,0.13}{##1}}}

\def\PYZbs{\char`\\}
\def\PYZus{\char`\_}

\def\PYZlt{\char`\<}
\def\PYZgt{\char`\>}
\def\PYZsh{\char`\#}

\def\PYZhy{\char`\-}
\def\PYZsq{\char`\'}
\def\PYZdq{\char`\"}

\makeatother

\journal{Astronomy and Computing}

\begin{document}

\begin{frontmatter}

\title{Comet: A VOEvent Broker}

\author{John Swinbank}
\ead{j.swinbank@uva.nl}

\address{Anton Pannekoek Institute, University of Amsterdam, Postbus 94249, 1090 GE Amsterdam, The Netherlands}

\begin{abstract}

The VOEvent standard provides a means of describing transient celestial events
in a machine-readable format. This is an essential step towards analysing and,
where appropriate, responding to the large volumes of transients which will be
detected by future large scale surveys. The VOEvent Transport Protocol (VTP)
defines a system by which VOEvents may be disseminated to the community. We
describe the design and implementation of Comet, a freely available, open
source implementation of VTP\@. We use Comet as a base to explore the
performance characteristics of the VTP system, in particular with reference to
meeting the requirements of future survey projects. We describe how, with the
aid of simple extensions to VTP, Comet can help users filter high-volume
streams of VOEvents to extract only those which are of relevance to particular
science cases.  Based on these tests and on the experience of developing
Comet, we derive a number of recommendations for future refinements of the VTP
standard.

\end{abstract}

\begin{keyword}
VOEvent \sep Astronomical transients \sep Time domain astrophysics \sep Network protocol design
\end{keyword}

\end{frontmatter}

\section{Introduction}
\label{sec:intro}

Exploring the astrophysical time domain through timely follow-up observations
of transient and variable sources offers the potential of many and varied
scientific results. However, achieving these results requires a fast and
reliable way of disseminating sufficient information about new transients to
appropriate follow-up facilities.

Mechanisms for distributing news of transient events already exist: both the
NASA Gamma-ray Coordinates Network\footnote{\url{http://gcn.gsfc.nasa.gov/}}
(GCN) and The Astronomer's
Telegram\footnote{\url{http://www.astronomerstelegram.org/}} have long track
records of enabling transient astronomy. However, the next generation of
large-scale survey telescopes such as Gaia, SKA and LSST promise an increase
by several orders of magnitude in the rate of transients being reported.  The
sheer volume of events presents a scalability challenge: it is no longer
practical for even large teams of astronomers to consider reading,
understanding and responding to these notifications manually. Automation is
essential. Furthermore, the diverse nature of these transient hunting
facilities---covering not just electromagnetic gamut from low-frequency radio
telescopes to space based X- and $\gamma$-ray monitors, but also other types
of instrumentation such as gravitational waves detectors---means that a
flexible and adaptable machine-readable mechanism must be adopted for
describing transients.

The International Virtual Observatory Alliance (IVOA) has developed the
VOEvent\footnote{\url{http://www.voevent.org/}} \citep{Seaman:2011} standard
to address these issues. VOEvent provides a standardized, machine- and
human-readable way of describing a wide range of transient astronomical
phenomena. An individual VOEvent document (or ``packet'') describes a
particular transient event, providing not only information about what has been
observed and how the observations were made, but also making it possible for
the author to include a scientific motivation for why this particular event is
interesting.  Furthermore, a VOEvent may cite other VOEvents, providing more
information about a given transient or, if necessary, superseding or
retracting an earlier message.

VOEvents are published as XML \citep{Bray:2008} documents which should be in
compliance with schema \citep{Gau:2012, Peterson:2012} produced by the IVOA\@.
Working in XML enables VOEvent to make extensive use of other relevant IVOA
standards and enables convenient processing with a wide range of commercial
and open-source software.

The VOEvent standard defines the structure and content of a VOEvent packet,
but it does not describe a mechanism by which the author of a VOEvent may
distribute it to potentially interested recipients. This transport agnosticism
is provides the maximum possible flexibility for individual projects to
disseminate events by whatever means best meets their science goals. However,
a baseline specification for a simple transport protocol is of value in terms
of providing a common starting point for building international VOEvent
distribution networks \citep{Williams:2012}.  The VOEvent Transport Protocol
\citep[VTP;][]{Allan:2009} is now seeing widespread adoption as such a
baseline.

This manuscript describes Comet, an implementation of all the components
necessary for interacting with VTP while acting as a test-bed for
production-level VTP deployments and for new technologies and ``value-added''
services to assist in addressing the transient deluge. Section~\ref{sec:vtp}
provides an overview of VTP and discusses the general topology of event
distribution networks. Section~\ref{sec:design} describes how Comet has been
designed and built to meet the protocol specifications.
Section~\ref{sec:filter} describes how Comet builds upon VTP to help address
future challenges in VOEvent filtering and selection. Section~\ref{sec:perf}
considers the performance implications of deploying VTP in support of
next-generation astronomical infrastructure, considering both the scalability
of the protocol to large numbers of events and to high latency connections. In
\S\ref{sec:security} we consider the security implications of VOEvents, how
they can be addressed at the transport level, and describe a system being
prototyped in Comet.  Implications for future revisions of the VTP standard
are summarized in \S\ref{sec:future}. Section~\ref{sec:avail} describes the
terms under which Comet is available and how to obtain it, while a summary of
the results are presented and some more general conclusions drawn for future
of VOEvent transport in \S\ref{sec:conclusions}.

\section{VOEvent Transport Protocol}
\label{sec:vtp}

\begin{figure}
  \begin{center}
  \includegraphics[width=\columnwidth]{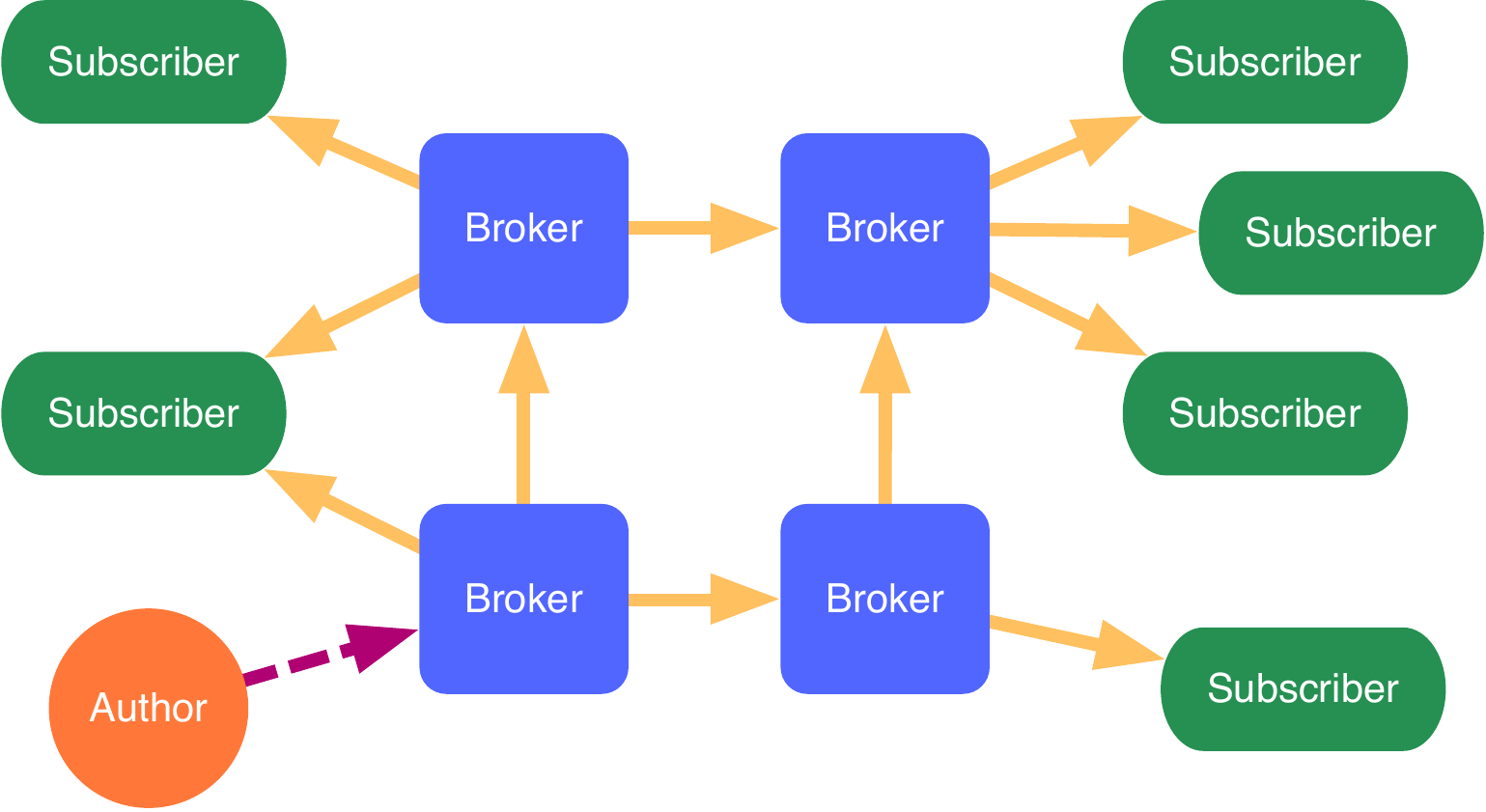}
  \end{center}

  \caption{An overview of the passage of a VOEvent through a VTP network. The
  network roles and connection types described in \S\ref{sec:vtp} are all
  represented. Arrows indicate the direction of data flow; their style
  represents the type of connection. The dashed arrow is an author-to-broker
  connection, while solid arrows are broker-to-subscriber connections
  (which are equivalent to broker-to-broker). First the event is sent by an
  author to a single broker. This broker then distributes it to all of its
  subscribers, which may include other brokers, which, in turn, redistribute
  the event until every entity on the network has received a single copy:
  de-duplication, where necessary, is applied as discussed in
  \S\ref{sec:design:dedup}. Adapted from \citet{Swinbank:2014}.}

  \label{fig:vtp}
\end{figure}

VTP provides a simple system for distributing VOEvents from one or more
authors to a network of potentially interested subscribers. It builds upon the
semantics of VOEvent interchange described in the VOEvent standard
\citep{Seaman:2011}, but includes only those entities which directly interact
by means of the network. To wit, VTP defines the following network roles:

\begin{description}

  \item[Author]{An author is responsible for creating and publishing one or
  more VOEvents.}

  \item[Subscriber]{A subscriber receives the VOEvents generated by one or
  more authors.}

  \item[Broker]{A broker receives VOEvents from other network entities
  re-distributes them to one or more subscribers. In addition, a broker may
  perform ``added value'' services. These could be at the request of
  particular subscribers (e.g.\ to apply a filter to the event stream sent
  to that subscriber), or applied more generally to the event stream (e.g.\ to
  apply some annotation to all events processed).}

\end{description}

Connections between these entities take place over TCP \citep{Cerf:1974}. The
VTP standard defines three types of connection:

\begin{description}

  \item[Author to Broker]{The author makes a TCP connection to the broker and
  transmits a VOEvent packet. On receipt of a syntactically valid message, the
  broker sends an acknowledgement.  The connection is then closed; submitting
  a further VOEvent packet would require initiating a new connection.}

  \item[Broker to Subscriber]{The subscriber opens a TCP connection to the
  broker, which remains open indefinitely. The broker and subscriber send
  periodic ``heartbeat'' messages over the connection to verify that it
  remains live.  When the broker receives an event for distribution, it sends
  it to the subscriber over this connection. The subscriber replies with an
  acknowledgement.}

  \item[Broker to Broker]{A broker may subscribe to the output of another
  broker. In doing so, it acts as a subscriber, and the relationship between
  them is as described in ``Broker to Subscriber'', above.}

\end{description}

Note that the broker-to-subscriber connection remains open at all times, even
when a subscriber has recently received an event. The standard mandates that
the subscriber must always be prepared to receive more events, even while a
previous event is still being processed: otherwise, a backlog of events
waiting to be sent to a particular subscriber could build up and overload the
broker.

By causing brokers to subscribe to the output of their peers, it is possible
to build extended networks of mutually-interconnected brokers. An author need
only publish to one broker and ultimately their event will be distributed to
all entities on the network. This is not only efficient, it is also robust:
the failure of any given entity can only cause local disruption to the
distribution system. The topology of such a network, and the path a VOEvent
packet might take across it, is shown in Fig.~\ref{fig:vtp}.

In addition to passing VOEvent XML documents, VTP defines a ``Transport''
document type. Transport documents are used for the heartbeat messages between
brokers and subscribers and for sending acknowledgement of event receipt. The
documents are kept intentionally short, providing simply a timestamp, an
indication of the originator, and---in the case of an acknowledgement---the
identity of the event being acknowledged.

VTP makes limited provision for securing access to the network: that is, for
limiting the authors and subscribers which may connect to a given broker. The
simplest, albeit least flexible, approach is for the broker to maintain a
``whitelist'' of the IP addresses of entities which are authorized to connect,
and simply drop connections coming from elsewhere. Such a system is convenient
and easy to implement for small networks, but can rapidly become unwieldy as
the list of authorized users grows or as those users need to connect from
multiple addresses. An alternative is therefore suggested in the standard
based on cryptographically signed transport messages, which enable an entity
to securely demonstrate its identity on connection. The means by which these
signatures may be applied is not specified in the VTP standard, which rather
refers to the systems proposed by \citet{Rixon:2005}, \citet{Denny:2008} and
\citet{Allen:2008}.  The application of cryptographic signatures to XML
documents is a potentially complex topic, and one to which we return in
\S\ref{sec:security}.

\section{The Design and implementation of Comet}
\label{sec:design}

Comet is a freely available, open source packages which can fulfil any or all
of the roles within a VTP network. It can receive events from remote brokers
(the subscriber role), receive events from authors and distribute them to
subscribers (the broker role) and it provides a tool which can publish a
VOEvent to a remote broker (the author role). Comet aims both to act as a
production-ready event distribution system, which projects can immediately
start using to service their science goals, and as a convenient system for
exploring the characteristics of VTP and prototyping future extensions to the
protocol. The first of these goals has already been achieved, with Comet
instrumental in low-latency follow up of gamma-ray bursts \citep{Staley:2013}.
Early results from the second goal are described in the subsequent sections of
this manuscript.

Version 1.1.0 of Comet was released in February of 2014 and is current stable
version at time of writing. Here, we specifically consider the implementation
of this version, although there are currently no plans for major architectural
changes in the future.

\subsection{Twisted Python and event-driven programming}
\label{sec:design:twisted}

Comet is implemented in Python, and is built atop the Twisted networking
engine\footnote{\url{https://twistedmatrix.com/}}. Twisted enables an
\textit{event-driven} and \textit{asynchronous} style of development which is
extensively used throughout Comet.

Conventionally, we think of programs as being executed in order: the system
executes the instructions described by the first statement, followed by the
second statement, and so on until the process is complete. Of course,
spreading a process across multiple threads of execution makes the precise
ordering of statements non-deterministic \citep[and, indeed, introduces a whole
new level of complexity in the process;][]{Lee:2006}, but the fundamental
point remains: the aim is to execute the program as rapidly and efficiently as
possible and then exit.

It is obvious that this model does not map well to network based applications.
Consider the ``subscriber'' role in a VOEvent network: it is not rushing to
finish some particular task and then terminate, but rather it continues
listening to the network indefinitely for the arrival of VOEvents, and takes
appropriate action when an packet is received. Event-driven programming is the
generalization of this concept: rather than a list of instructions to be
executed sequentially, we define the actions that should be taken in response
to possible events. Twisted provides an ``event loop'' which waits for events
and calls the appropriate actions when they occur.

\begin{listing}[t]
\hrulefill
\vspace{-10pt}
\begin{Verbatim}[commandchars=\\\{\}]
\PY{k}{class} \PY{n+nc}{VOEventReceiver}\PY{p}{(}\PY{n}{Protocol}\PY{p}{)}\PY{p}{:}
  \PY{n}{TIMEOUT} \PY{o}{=} \PY{l+m+mi}{20} \PY{c}{\PYZsh{} seconds}

  \PY{k}{def} \PY{n+nf}{connectionMade}\PY{p}{(}\PY{n+nb+bp}{self}\PY{p}{)}\PY{p}{:}
    \PY{n}{setTimeout}\PY{p}{(}\PY{n+nb+bp}{self}\PY{o}{.}\PY{n}{TIMEOUT}\PY{p}{)}

  \PY{k}{def} \PY{n+nf}{connectionLost}\PY{p}{(}\PY{n+nb+bp}{self}\PY{p}{)}\PY{p}{:}
    \PY{n}{setTimeout}\PY{p}{(}\PY{n+nb+bp}{None}\PY{p}{)}
    \PY{n}{close\PYZus{}connection}\PY{p}{(}\PY{p}{)}

  \PY{k}{def} \PY{n+nf}{timeoutConnection}\PY{p}{(}\PY{n+nb+bp}{self}\PY{p}{)}\PY{p}{:}
    \PY{n}{log}\PY{o}{.}\PY{n}{msg}\PY{p}{(}\PY{l+s}{\PYZdq{}}\PY{l+s}{Connection timed out}\PY{l+s}{\PYZdq{}}\PY{p}{)}
    \PY{n}{close\PYZus{}connection}\PY{p}{(}\PY{p}{)}

  \PY{k}{def} \PY{n+nf}{stringReceived}\PY{p}{(}\PY{n+nb+bp}{self}\PY{p}{,} \PY{n}{data}\PY{p}{)}\PY{p}{:}
    \PY{k}{try}\PY{p}{:}
        \PY{n}{message} \PY{o}{=} \PY{n}{parse}\PY{p}{(}\PY{n}{data}\PY{p}{)}
        \PY{k}{if} \PY{n}{is\PYZus{}valid}\PY{p}{(}\PY{n}{message}\PY{p}{)}\PY{p}{:}
            \PY{n}{log}\PY{o}{.}\PY{n}{info}\PY{p}{(}\PY{l+s}{\PYZdq{}}\PY{l+s}{Good message received}\PY{l+s}{\PYZdq{}}\PY{p}{)}
            \PY{n}{acknowledge}\PY{p}{(}\PY{n}{message}\PY{p}{)}
            \PY{n}{process\PYZus{}event}\PY{p}{(}\PY{n}{message}\PY{p}{)}
        \PY{k}{else}\PY{p}{:}
            \PY{n}{log}\PY{o}{.}\PY{n}{warning}\PY{p}{(}\PY{l+s}{\PYZdq{}}\PY{l+s}{Bad message received}\PY{l+s}{\PYZdq{}}\PY{p}{)}
    \PY{k}{except} \PY{n}{ParseError}\PY{p}{:}
        \PY{n}{log}\PY{o}{.}\PY{n}{warning}\PY{p}{(}\PY{l+s}{\PYZdq{}}\PY{l+s}{Message unparsable}\PY{l+s}{\PYZdq{}}\PY{p}{)}
    \PY{k}{finally}\PY{p}{:}
        \PY{n}{close\PYZus{}connection}\PY{p}{(}\PY{p}{)}
\end{Verbatim}

\vspace{-15pt}
\hrulefill
\caption{An example of an event-driven Twisted protocol, based on Comet's
\texttt{VOEventReceiver}.}
\label{lst:event}
\end{listing}

When talking to the network, Twisted provides the \textit{Protocol} as an
abstraction for managing events. A protocol defines the interaction that a
particular component of the system has with the network. For example,
Listing~\ref{lst:event} shows a simplified version of the protocol for Comet's
\texttt{VOEventReceiver}. This is the part of the broker which listens to the
network for submissions from authors. Four separate events are handled by this
protocol:

\begin{itemize}

\item{When a new connection is initiated by an author, the broker sets a timer
on the connection. If no traffic is received, the timer will eventually reach
zero and the connection will be timed-out. The timer is initialized to the
essentially arbitrary value of 20\,s; this may be refined (or made
user-configurable) in light of practical experience in future Comet releases.}

\item{When a connection is lost, the connection is closed and the timeout is
aborted.}

\item{When a connection times out, close it.}

\item{When a string is received over the connection, parse it and see if it
can be recognized as a valid VOEvent. If so, return an acknowledgement and
process the newly received event (for example by re-distributing it to
subscribers). If not, log a warning message. Finally, shut down the
connection.}

\end{itemize}

Similar, although often more complex, protocols are defined for all of the
other roles in the system: an author connecting to a broker
(\texttt{VOEventSender}), a broker to a subscriber
(\texttt{VOEventBroadcaster}), and an subscriber to a broker
(\texttt{VOEventSubscriber}).

Event-driven programming provides a convenient abstraction for responding to
network events. However, it does not address issues regarding concurrency. As
described in \S\ref{sec:vtp}, VTP requires that even immediately after
receiving an event subscribers must be ready to accept another: there can be
no delay while the event is ingested. Contrast this with the model described
above and outlined in Listing~\ref{lst:event}: here, when an event is
received, each of the functions \texttt{parse()}, \texttt{is\_valid()},
\texttt{acknowledge()}, \texttt{process\_event()} and
\texttt{close\_connection()} is called in turn.  If these operations are not
assumed to be instantaneous, we must wait for them to complete before
proceeding. While waiting, new events cannot be received.  We are thus in
violation of the VTP standard\footnote{In practice, the implementation of some
of these operations used in the Comet codebase can be assumed to be
effectively instantaneous.  This is safe so long as the time taken to parse is
sufficiently short that no backlog of events waiting to be processed  builds
up and no network timeouts occur.}.

Twisted addresses this problem through the use of \textit{Deferred}s. A
deferred is effectively a promise that processing is underway and that results
will be available in future. We can then queue up other processing tasks (or
``callbacks'') that will be executed when the result of the deferred is
available. For example, we could define a version of
\texttt{parse()}---call it
\texttt{deferred\_parse()}---that, rather than returning an object
representing a parsed version of the VOEvent document, returns a promise to
eventually parse the document in the future and then make it available for
further processing. We can then queue up our other functions to run only when
parsing is complete. For example, see Listing~\ref{lst:deferred}, in which we
queue up a number of callbacks to be run when parsing is complete and also add
an ``errback'' which handles logging a message if any of the callbacks fail to
run successfully.

\begin{listing}[H]
\hrulefill
\vspace{-10pt}
\begin{Verbatim}[commandchars=\\\{\}]
\PY{k}{def} \PY{n+nf}{stringReceived}\PY{p}{(}\PY{n+nb+bp}{self}\PY{p}{,} \PY{n}{data}\PY{p}{)}\PY{p}{:}
  \PY{n}{d} \PY{o}{=} \PY{n}{deferred\PYZus{}parse}\PY{p}{(}\PY{n}{data}\PY{p}{)}
  \PY{n}{d}\PY{o}{.}\PY{n}{addCallback}\PY{p}{(}\PY{n}{is\PYZus{}valid}\PY{p}{)}
  \PY{n}{d}\PY{o}{.}\PY{n}{addCallback}\PY{p}{(}\PY{n}{check\PYZus{}role}\PY{p}{)}
  \PY{n}{d}\PY{o}{.}\PY{n}{addCallback}\PY{p}{(}\PY{n}{acknowledge}\PY{p}{)}
  \PY{n}{d}\PY{o}{.}\PY{n}{addCallback}\PY{p}{(}\PY{n}{process\PYZus{}event}\PY{p}{)}
  \PY{n}{d}\PY{o}{.}\PY{n}{addErrback}\PY{p}{(}\PY{n}{log\PYZus{}failure}\PY{p}{)}
  \PY{n}{d}\PY{o}{.}\PY{n}{addCallback}\PY{p}{(}\PY{n}{close\PYZus{}connection}\PY{p}{)}
\end{Verbatim}

\vspace{-15pt}
\hrulefill
\caption{A version of \texttt{VOEventReceiver.stringReceived()} (shown in
Listing~\ref{lst:event}) based on deferred processing.}
\label{lst:deferred}
\end{listing}

Finally, we must implement \texttt{deferred\_parse()}.  Simply returning a
deferred from a function does not prevent it from blocking.  Instead, we
create a dedicated thread which is devoted to parsing the data, and have it
run concurrently with the rest of the application. When that thread completes,
the deferred fires with its result. Conveniently, Twisted makes it easy to
apply this pattern to a blocking function such as our \texttt{parse()}: see
Listing~\ref{lst:deferToThread}.

\begin{listing}[H]
\hrulefill
\vspace{-10pt}
\begin{Verbatim}[commandchars=\\\{\}]
\PY{k+kn}{from} \PY{n+nn}{twisted.internet.threads} \PY{k+kn}{import} \PYZbs{}
     \PY{n}{deferToThread}

\PY{k}{def} \PY{n+nf}{deferred\PYZus{}parse}\PY{p}{(}\PY{n}{data}\PY{p}{)}\PY{p}{:}
    \PY{k}{return} \PY{n}{deferToThread}\PY{p}{(}\PY{n}{parse}\PY{p}{,} \PY{n}{data}\PY{p}{)}
\end{Verbatim}

\vspace{-15pt}
\hrulefill
\caption{The implementation of the non-blocking \texttt{deferred\_parse()}
function.}
\label{lst:deferToThread}
\end{listing}

Although the examples presented in this section are only intended to be
illustrative, they demonstrate the concepts of asynchronous, event-driven
programming upon which Comet is built and are fundamental to understanding its
operation.

It is worth emphasizing that the techniques described in this section are not
unique to Twisted. Other frameworks such as
gevent\footnote{\url{http://www.gevent.org/}} and asyncio\footnote{Recently
added to the Python standard library;
\url{https://docs.python.org/3.4/library/asyncio.html}} provide
implementations of similar capabilities in Python, and equivalent libraries
are available for many other languages. However, the rich software ecosystem
supported by Twisted, combined with its demonstrated ability to deliver
acceptable performance (\S\ref{sec:perf}), have provided an excellent platform
upon which to develop Comet.

\subsection{Comet architecture}
\label{sec:design:architecture}

Comet is built around the four Twisted protocols discussed in
\S\ref{sec:design:twisted}. These enable it to take the part of either side in
each of the three connection types discussion in \S\ref{sec:vtp}. For the
convenience of end users, these are made accessible under two distinct front
ends. In this section, we first introduce those components and describe the
relationship between them, and then discuss how Comet implements some specific
requirements of VTP.

\subsubsection{The components of Comet}
\label{sec:design:components}

The authorial component is \textit{comet-sendvo}. This is a command-line tool
which enables the user to submit a VOEvent to a remote broker. The user is
expected to supply the VOEvent either on standard input or via a reference to
the filesystem; \textit{comet-sendvo} transmits it to the specified
destination using the \texttt{VOEventSender} protocol and shuts down.

Processing a single event and then exiting is an appropriate model for an
author, but is not the behaviour required of a broker or subscriber. Rather,
these tools must remain active, continuing to receive and process VOEvents
until the user shuts them down. To support this mode of operation, Comet can
run as a ``daemon'', or background process. The Comet daemon can:

\begin{enumerate}

  \item{Accept submissions from authors (including, of course, \textit{comet-sendvo});}

  \item{Subscribe to event streams from one or more remote brokers;}

  \item{Distribute event packets received (whether by direct author submission or by subscription) to its own subscribers;}

  \item{Execute arbitrary logic based upon the event packets received.}

\end{enumerate}

A single Comet daemon is capable of performing any or all of these actions,
depending upon configuration: it is not necessary to start separate ``broker''
and ``subscriber'' daemons, for example.

Both \textit{comet-sendvo} and the Comet daemon make extensive use of the
facilities provided by Twisted for event-driven and asynchronous programming,
as well as its support for logging and daemonization. They are exclusively
command-line driven, and do not rely on configuration files.

\subsubsection{Schema validation}
\label{sec:design:schema}

It is possible to construct XML documents which claim to be VOEvents but which
do not, in fact, adhere to the VOEvent XML schema. In some cases, the document
may be completely unparsable; in others, it may be possible to extract some
data, but with unpredictable results and no guarantee that the recipient
receives the information the author intended.

Current versions of Comet insists that events being submitted to the broker by
an author comply with the VOEvent 2.0
schema\footnote{\url{http://www.ivoa.net/xml/VOEvent/VOEvent-v2.0.xsd}};
extending this to include later versions as they become available is
straightforward. Schema validation is applied to the event before it is
accepted for redistribution by the broker; if that validation fails, a
\texttt{nak} message which indicates the problem is sent to the author and the
event is dropped. It is to be hoped that a well-intentioned author will
correct and resubmit the event.

When receiving events from an upstream broker (either as a broker itself or as
a subscriber) Comet does not attempt to validate the event against the schema
(although it is still required that the event must be parsable). This is
because there is no way to indicate the failure to the author: any
\texttt{nak} sent to the upstream broker will not be propagated further. The
author cannot know that their event has been rejected, will not correct and
re-send, and valuable scientific content may be lost.

\subsubsection{Event de-duplication}
\label{sec:design:dedup}

As described in \S\ref{sec:vtp} it is possible to build a mutually
interconnected ``mesh'' of brokers to efficiently and reliably distribute
VOEvent packets to a large number of subscribers. However, this runs the risk
that events could continue ``looping'' on the network indefinitely, as two or
more brokers which subscribe to each other's feeds repeatedly exchange the
same event. To avoid this problem, Comet refuses to process any given event
more than once: if a newly received event is the same as one which has been
previously seen, it is simply dropped without being further distributed.

In order for this approach to be effective, it is necessary to define what it
means for two VOEvent documents to be ``the same''. In particular, due to the
nature of XML, it is possible for exactly the same information about an event
(the ``infoset'') to be encoded in multiple different, but all equally valid,
XML documents. At the simplest level, this is because XML is (for the most
part) white space agnostic---new lines or spaces can be inserted without
changing the meaning of the document. The question becomes more complex,
though, when we consider the various versions of the VOEvent standard. Version
2.0 \citep{Seaman:2011} is current, but the previous version
\citep[1.1;][]{Seaman:2006} is still in use by some systems. If the same
information about the same astronomical event is encoded in a VOEvent 2.0
document and a VOEvent 1.11 document, are these ``the same''?

This question is particularly pertinent because this situation is exactly that
which exists in practice: since
2012\footnote{\url{http://gcn.gsfc.nasa.gov/admin/voevent_version20_available.txt}},
NASA GCN has issued both version 1.1 and version 2.0 VOEvents containing the
same information.

\citet{Seaman:2011} requires that each VOEvent carry an
IVORN\footnote{International Virtual Observatory Resource Name} which ``will
stand in for a particular packet''. It is this IVORN that is used to identify
events in the context of references and citations, for example. There has been
some
debate\footnote{\url{http://www.ivoa.net/pipermail/voevent/2012-March/002836.html}}
as to whether IVORNs uniquely identify a particular infoset or a particular
representation thereof: this is not currently well defined by the relevant
documentation. In the case of the events issued by GCN, a single IVORN is used
to describe both the version 1.1 and the version 2.0 VOEvent packets: this
provide a \textit{de facto} standard that the IVORN identifies the infoset.

VTP makes no distinction between the versions of VOEvents which it transmits:
the same protocol may be used for version 1.1, or 2.0, or putative future
versions. However, the consumer of a particular VOEvent may well have a
toolchain that is tuned to work with one particular standard. In other words,
authors may wish to use a VOEvent network to distribute multiple different
versions of the same event, while subscribers may depend on receiving a
specific representation of that event. If an author submits version 1.1 and
version 2.0 representations of an event, it would not be appropriate for a
broker to regard them as duplicates and discard one of them. The only possible
conclusion is that the IVORN is not a suitable means of identifying unique
packets for the purposes of de-duplication.

Comet therefore regards packets as duplicates only if they are bit-for-bit
identical with an packet which has been seen before. This is determined by
calculating the SHA-1 \citep{Eastlake:2001} cryptographic hash of every packet
which is seen by a Comet daemon and storing it, together with the time and
date at which the packet was seen, in a DBM-style \citep{DBM:1979} persistent
database\footnote{``DBM-style'' databases provide mappings between ``keys''
and ``values'' in the manner of an associative array. Various libraries
implementing this style of database exist; Comet uses Python's \texttt{anydbm}
interface, which automatically chooses a particular implementation based upon
the platform on which it is running.}. When an event is received, its SHA-1
hash is calculated and compared against the contents of the database to
establish if it has been seen before.

Each individual SHA-1 hash is stored as 40 bytes, plus a further 13 bytes are
used to records the timestamp. The total storage requirement is therefore very
modest. However, on a busy broker processing many events, the database could
grow to a significant size, wasting resources and slowing down access.
Therefore, Comet periodically removes all events older than 30 days from its
database. Duplicates issued more than 30 days after the original event will
therefore not be detected; however, an event loop with such a long period
poses no threat to the integrity of the network. It may be appropriate to tune
this timescale in future based on practical experience with large scale VTP
deployments.

It should be noted that this de-duplication scheme requires that all entities
on the network forward events unchanged: even an apparently inconsequential
change to an event packet which results in a valid encoding the same infoset
as before would result in a different SHA-1 hash for the event. The current
VTP standard implies but does not absolutely require this behaviour: see
\S\ref{sec:future} for further discussion.

\subsubsection{Security and whitelisting}
\label{sec:design:security}

The released version of Comet described here (1.1.0; \S\ref{sec:design}) does
not implement an authentication scheme based on cryptographic signatures as
described in \S\ref{sec:vtp}. Work is ongoing on prototyping such a scheme
using Comet as a test-bed: this is described in \S\ref{sec:security}.

When acting as a broker, Comet includes the ability to check authors
submitting events against a whitelist of IP addresses. Multiple disjoint
ranges of addresses to whitelist may be specified using CIDR notation
\citep{Fuller:1993}, making the system very flexible.

Comet does not currently provide built-in whitelisting support for
subscribers. However, equivalent functionality is available though the use of
an operating system level packet filter.

\subsubsection{Acting on events received}
\label{sec:design:plugin}

Just receiving a VOEvent and optionally re-distributing it is of limited
practical value: ultimately, some recipient of the event will wish to take
action based upon it. The algorithms which may be employed to determine
whether a given event is worth of follow-up are dependent on the particular
science goals of the recipient, are potentially complex, and are certainly
outside the scope of this manuscript. Since it is not possible to anticipate
the requirements of the end user in a universally applicable way, Comet rather
seeks to be easily adapted to each particular use case. Two mechanisms are
provided to make this possible.

The simplest option is that when a new event is received, Comet can spawn an
external process and provide the text of the event packet to it on standard
input.  The process is run asynchronously, so that potentially lengthy
processing jobs can be run on events without interrupting Comet's regular
operation.  Comet monitors the execution of the process and logs a warning if
it is unsuccessful (that is, if it exits with a status other than \texttt{0}),
but otherwise has no control over the processing performed.

In some circumstances, the user may wish for more control than is provided for
by passing events to another process. Comet therefore makes it possible to
write \textit{plugins}, which can be loaded into the daemon at run time.
Users write plugins in Python, implementing a standard interface. Plugins
provide a \texttt{\_\_call\_\_()} method which is invoked with the contents of
an event whenever one is received. An example is shown in
Listing~\ref{lst:plugin}.

\begin{listing}[t]
\hrulefill
\vspace{-10pt}
\begin{Verbatim}[commandchars=\\\{\}]
\PY{k+kn}{from} \PY{n+nn}{zope.interface} \PY{k+kn}{import} \PY{n}{implementer}
\PY{k+kn}{from} \PY{n+nn}{twisted.plugin} \PY{k+kn}{import} \PY{n}{IPlugin}
\PY{k+kn}{from} \PY{n+nn}{comet.icomet} \PY{k+kn}{import} \PY{n}{IHandler}\PY{p}{,} \PY{n}{IHasOptions}

\PY{c}{\PYZsh{} A plugin implements the IPlugin}
\PY{c}{\PYZsh{} and IHandler interfaces}
\PY{n+nd}{@implementer}\PY{p}{(}\PY{n}{IPlugin}\PY{p}{,} \PY{n}{IHandler}\PY{p}{)}
\PY{k}{class} \PY{n+nc}{ExamplePlugin}\PY{p}{(}\PY{n+nb}{object}\PY{p}{)}\PY{p}{:}
    \PY{c}{\PYZsh{} The \PYZdq{}name\PYZdq{} attribute is used to refer}
    \PY{c}{\PYZsh{} to the plugin on the command line.}
    \PY{n}{name} \PY{o}{=} \PY{l+s}{\PYZdq{}}\PY{l+s}{example}\PY{l+s}{\PYZdq{}}

    \PY{c}{\PYZsh{} The \PYZdq{}\PYZus{}\PYZus{}call\PYZus{}\PYZus{}()\PYZdq{} method is invoked}
    \PY{c}{\PYZsh{} when a new event is received.}
    \PY{k}{def} \PY{n+nf}{\PYZus{}\PYZus{}call\PYZus{}\PYZus{}}\PY{p}{(}\PY{n+nb+bp}{self}\PY{p}{,} \PY{n}{event}\PY{p}{)}\PY{p}{:}
        \PY{k}{print} \PY{l+s}{\PYZdq{}}\PY{l+s}{Event received}\PY{l+s}{\PYZdq{}}

\PY{c}{\PYZsh{} The plugin must be instantiated before use.}
\PY{n}{example\PYZus{}plugin} \PY{o}{=} \PY{n}{ExamplePlugin}\PY{p}{(}\PY{p}{)}
\end{Verbatim}

\vspace{-15pt}
\hrulefill
\caption{A simple example of a Comet event handling plugin. This plugin prints
a message whenever a new event is received.}
\label{lst:plugin}
\end{listing}

Comet automatically probes for all available plugins and makes them available
as command line arguments, so the user can specify which plugins are required
when the daemon is started. If required, plugins may also define configuration
parameters implementing the \texttt{IHasOptions} interface; these are exposed
as command line options.

\section{Filtering events}
\label{sec:filter}

Next-generation telescopes such as LSST anticipate to detecting and announcing
transients using VOEvent at rates of perhaps tens of millions of events per
day \citep{Kantor:2014}. It is unlikely that most individual subscribers will
have a use for all of these events. Winnowing that event stream down so that
each subscriber receives only those events which are of direct relevance to
them is both efficient in terms of resource usage, as fewer events need to be
transported to and processed by the subscriber, but also enables subscribers
to deploy simple, well-targeted systems that address their science goals,
rather than attempting to devise efficient ways to process millions of VOEvent
packets.

Efforts to develop intelligent systems for alerting users only of those events
which are of relevance to them are ongoing, and will continue into the future
\citep{Williams:2009}. Comet contributes to this effort by introducing a
powerful XPath \citep{Clark:1999} based filtering system.

\subsection{XPath queries}
\label{sec:filter:xpath}

XPath is a language for selecting parts of and computing values over an XML
document. XPath expressions may return one of four different result types:

\begin{itemize}
  \item{A Boolean value;}
  \item{A floating point number;}
  \item{A textual string;}
  \item{A set of XML ``nodes'', representing parts of the document.}
\end{itemize}

XPath enables users to specify complex queries, including testing
the values of arbitrary elements or attributes specified in the document and
combining those tests with Boolean logic. A complete reference is outside the
scope of this manuscript, but some examples may serve to illustrate the
possibilities.

Starting with string matching,

\begin{xpath}[H]
\begin{Verbatim}[commandchars=\\\{\}]
//Who/Author[shortName=\PYZdq{}VO\PYZhy{}GCN\PYZdq{}]
\end{Verbatim}

\vspace{-15pt}
\hrulefill
\vspace{-9pt}
\captionsetup{labelsep=none,textformat=empty,justification=raggedleft,singlelinecheck=false}
\caption{Not shown}
\label{xpath:exp1}
\end{xpath}

returns a set of all nodes in the document which list the author's ``short
name'' as \texttt{VO-GCN}. More complex matches can use functions, such as

\begin{xpath}[H]
\begin{Verbatim}[commandchars=\\\{\}]
//How[contains(Description, \PYZdq{}Swift\PYZdq{})]
\end{Verbatim}

\vspace{-15pt}
\hrulefill
\vspace{-9pt}
\captionsetup{labelsep=none,textformat=empty,justification=raggedleft,singlelinecheck=false}
\caption{Not shown}
\label{xpath:exp2}
\end{xpath}

which returns the set of all nodes which mention ``Swift'' in the context of a
describing how the data was obtained. Numerical comparisons are also possible:

\begin{xpath}[H]
\begin{Verbatim}[commandchars=\\\{\}]
//Param[@name=\PYZdq{}Sun\PYZus{}Distance\PYZdq{} and @value\PYZgt{}40]
\end{Verbatim}

\vspace{-15pt}
\hrulefill
\vspace{-9pt}
\captionsetup{labelsep=none,textformat=empty,justification=raggedleft,singlelinecheck=false}
\caption{Not shown}
\label{xpath:exp3}
\end{xpath}

provides the set of all parameters called \texttt{Sun\_Distance} with a
numerical value greater than 40.

These expression can be combined, so that for example

\begin{xpath}[H]
\begin{Verbatim}[commandchars=\\\{\}]
//How[contains(Description, \PYZdq{}Swift\PYZdq{})] or
  ( //Param[@name=\PYZdq{}Sun\PYZus{}Distance\PYZdq{} and @value\PYZgt{}40]
    and //Who/Author[shortName=\PYZdq{}VO\PYZhy{}GCN\PYZdq{}]) )
\end{Verbatim}

\vspace{-15pt}
\hrulefill
\vspace{-9pt}
\captionsetup{labelsep=none,textformat=empty,justification=raggedleft,singlelinecheck=false}
\caption{Not shown}
\label{xpath:exp4}
\end{xpath}

returns a Boolean value which is true if the event either mentions ``Swift''
or both has a \texttt{Sun\_Distance} parameter greater than 40 and originates
from GCN, and false otherwise.

\subsection{Integration with Comet}

Comet makes it possible for a subscriber to supply one or more XPath
expressions to a broker. When the broker receives an event, it evaluates each
expression over the event, and only forward it to the subscriber if at least
one of the expressions evaluates produces a positive result.

Comet takes the result returned by XPath and applies Python's \texttt{bool()}
built-in function to determine if the result is ``positive''. For example, the
values \texttt{True}, \texttt{1} and \texttt{"string"} (the Boolean true
value, a non-zero number and a non-empty string) as well as a non-empty node
set are all positive, while \texttt{False}, \texttt{0} and \texttt{""}
(Boolean false, the number 0 and an empty string) and the empty node set are
``negative'' results.

VTP provides no standardized method for a subscriber to send their filter
preferences to the broker. Comet works around this by overloading the
Transport message system provided by VTP and described in \S\ref{sec:vtp}.
According to the VTP specification, it is legal for a subscriber to send a
Transport message of class \texttt{authenticationresponse} and with arbitrary
metadata embedded at any time during a VTP session. Comet looks for XPath
expressions encoded in this metadata and installs them as filters for the
subscriber; other brokers which comply with the protocol but do not support
this form of filtering should simply ignore the message.

\subsection{Alternative filtering systems}

XPath provides a convenient, standardized and expressive language for
accessing and performing simple calculations and comparisons based upon the
contents of XML documents. Incorporating XPath based filtering into Comet was
straightforward and doing so provides a powerful means of winnowing
high-volume VOEvent streams.

However, XPath is not appropriate for meeting every possible use case. In
particular, XPath expressions are evaluated over individual VOEvents, with no
reference to their surrounding context. Consequently, XPath expressions cannot
be used to draw scientific conclusions---or even perform rate-limiting---based
on the evolving contents of a stream of events. Further, XPath provides no
specialist astronomical or mathematical routines: it is impractical to use it
for filtering based on operations beyond simple arithmetic and comparisons.

Given these considerations, it is likely that addressing some scientific goals
will require a different approach to filtering than that currently supported
by Comet. The VTP system explicitly allows for this by encouraging brokers to
layer arbitrary ``added value'' services on top of the basic VTP system: a
richer, more astronomically-focused and context-aware filtering system is an
example of the possibilities. Indeed, such a service has precedent in the form
of SkyAlert \citep{Williams:2009}, which provides a Python-based interface to
filtering events.

\section{Performance}
\label{sec:perf}

Comet not been designed primarily for performance: at time of writing, typical
VOEvent brokers are processing perhaps a few hundred events per day, so the
total computational and storage demands are extremely modest. However, it is
informative to consider both how Comet and the VTP architecture scale to cope
with the millions of events per night promised by future facilities such as
LSST\@. In this section, we quantify both the number of events Comet is
capable of processing, the latency which it introduces to the event stream,
and the number of subscribers which a broker can conveniently service. We
begin by describing the test system, move on to discuss the performance
characteristics of the major operations which Comet performs when processing
an individual VOEvent message, and then take a more holistic approach to
consider the performance of a networked Comet broker under a variety of loads.

\subsection{Test system configuration}
\label{sec:perf:system}

The basic configuration of all tests below consists of one or more authors
connecting to a broker and sending it events which the broker then distributes
to one or more subscribers. The processes acting as authors, brokers and
subscribers were all run on the same modest desktop system, based on an Intel
Core i7 940 CPU\footnote{Four cores with two threads each running at
2.93\,GHz.} and 8\,GiB RAM\@. Storage was provided by two 7200\,RPM magnetic
disks configured as a RAID-0 array. The system was running
Debian\footnote{\url{http://www.debian.org/}} GNU/Linux with kernel version
3.13.

In realistic scenarios, VOEvent authors, brokers and subscribers would not
co-exist on the same system. However, providing many separate test systems was
impractical, and exchanging events over the public internet (or even over a
local network) introduces an extra layer of uncertainty in terms of network
latency. Instead, the various processes being tested were run in isolated
process containers using Docker\footnote{\url{https://www.docker.io/}}.
Docker-based containers operate in much the same way as traditional virtual
machines, except that they incur no virtualization overhead. They directly
address the same kernel as the host system, but are only able to communicate
with each other over (virtual) network interfaces. Within each container a
minimal Ubuntu\footnote{\url{http://www.ubuntu.com/}} Linux 12.04 system was
installed, providing Python 2.7.3 and Twisted 11.1.0. All testing was carried
out with Comet 1.1.0. The ``Dockerfile'' used to create exactly the system
used for these tests, as well as all the benchmarking scripts and plugins
described below, are available from the Comet repository (\S\ref{sec:avail}).

\subsection{Individual event processing}
\label{sec:perf:individual}

When an event is received from an author by the Comet broker for redistribution to
subscribers it passes through five distinct processing stages. These are:

\begin{enumerate}

  \item{The XML document text is parsed into an internal data structure;}

  \item{The VOEvent is checked for validity against the VOEvent 2.0 XML schema
  (\S\ref{sec:design:schema});}

  \item{The SHA-1 hash of the document text is calculated;}

  \item{The hash is compared against, and, if necessary, appended to the
  database of previously seen VOEvents (\S\ref{sec:design:dedup});}

  \item{Optionally, one or more XPath expressions are evaluated against the
  document before it is forwarded to each subscriber (\S\ref{sec:filter}).}

\end{enumerate}

Most of these operations are likely to depend upon the particular VOEvent
document being handled: a longer and more complex message will naturally
require more effort to process (the exception is checking and recording the
document against the event database, which involves processing just the IVORN
rather than the complete document). To best represent a real-world workload,
the tests were carried out using a corpus of 16425 VOEvents harvested from
currently operation VTP brokers between 5 and 15 July 2014\footnote{All
documents claiming to comply with the VOEvent 2.0 schema which were
distributed by any of \url{voevent.phys.soton.ac.uk}, \url{voevent.dc3.com},
\url{voevent.swinbank.org}, \url{68.169.57.253}, \url{209.208.78.170} or
\url{50.116.49.68} were collected. The three numerical IPv4 addresses are used
by NASA GCN and do not have DNS PTR records.}. The VOEvents originated from a
variety of sources, and include both notifications of astronomical phenomena
and sundry utility and test messages. The longest document consisted of 9647
bytes; the shortest 636; the median length was 5002 bytes.

Sections~\ref{sec:perf:individual:parse}--\ref{sec:perf:individual:xpath},
below, describe tests carried out to investigate the performance of each of
these operations in turn. A summary of the results is presented in
\S\ref{sec:perf:individual:results}.

\subsubsection{XML parsing}
\label{sec:perf:individual:parse}

All 16425 VOEvent documents in the test corpus were read from disk and stored
as textual data in memory. Each in turn was parsed into Comet's internal
VOEvent representation\footnote{Comet represents XML documents using a
custom-built wrapper around the \texttt{Element} class provided by lxml
(\url{http://lxml.de/}).}. In order to confirm that parsing was successful,
the Comet API was used to retrieve the \texttt{version} attribute from the
parsed document and confirm that it was equal to \texttt{"2.0"}. The total
time taken to parse and read the attribute from all of the events was
measured.

\subsubsection{Schema validation}

All test VOEvent documents were read from disk, parsed, and stored in memory
using Comet's internal representation. The VOEvent 2.0 XML schema was also
read from disk and parsed into an lxml \texttt{XMLSchema} object, the same
data structure as used by Comet for schema validation during normal
operations. The total time taken to check all the events against the schema
was measured. Two of the events failed validation.

\subsubsection{SHA-1 calculation}
\label{sec:perf:individual:hash}

All VOEvent documents in the test corpus were read from disk, parsed, and
stored in memory using Comet's internal representation. The total time taken
to calculate the 40 byte hexadecimal SHA-1 hash for each event in turn was
measured.

\subsubsection{Event database operations}
\label{sec:perf:individual:eventdb}

\begin{listing*}
\hrulefill
\vspace{-10pt}
\begin{Verbatim}[commandchars=\\\{\}]
\PY{c+cp}{\PYZlt{}?xml version=\PYZsq{}1.0\PYZsq{} encoding=\PYZsq{}UTF\PYZhy{}8\PYZsq{}?\PYZgt{}}
\PY{n+nt}{\PYZlt{}voe:VOEvent}
  \PY{n+na}{xmlns:voe=}\PY{l+s}{\PYZdq{}http://www.ivoa.net/xml/VOEvent/v2.0\PYZdq{}}
  \PY{n+na}{xmlns:xsi=}\PY{l+s}{\PYZdq{}http://www.w3.org/2001/XMLSchema\PYZhy{}instance\PYZdq{}}
  \PY{n+na}{ivorn=}\PY{l+s}{\PYZdq{}ivo://comet.broker/test\PYZsh{}TestEvent\PYZhy{}2014\PYZhy{}03\PYZhy{}31T16:16:37\PYZdq{}} \PY{n+na}{role=}\PY{l+s}{\PYZdq{}test\PYZdq{}} \PY{n+na}{version=}\PY{l+s}{\PYZdq{}2.0\PYZdq{}}
  \PY{n+na}{xsi:schemaLocation=}\PY{l+s}{\PYZdq{}http://www.ivoa.net/xml/VOEvent/v2.0}
\PY{l+s}{    http://www.ivoa.net/xml/VOEvent/VOEvent\PYZhy{}v2.0.xsd\PYZdq{}}
  \PY{n+nt}{\PYZgt{}}
\PY{n+nt}{\PYZlt{}Who}\PY{n+nt}{\PYZgt{}}
  \PY{n+nt}{\PYZlt{}AuthorIVORN}\PY{n+nt}{\PYZgt{}}ivo://comet.broker/test\PY{n+nt}{\PYZlt{}/AuthorIVORN\PYZgt{}}
  \PY{n+nt}{\PYZlt{}Date}\PY{n+nt}{\PYZgt{}}2014\PYZhy{}03\PYZhy{}31T16:16:37.040340\PY{n+nt}{\PYZlt{}/Date\PYZgt{}}
\PY{n+nt}{\PYZlt{}/Who\PYZgt{}}
  \PY{n+nt}{\PYZlt{}What}\PY{n+nt}{\PYZgt{}}
    \PY{n+nt}{\PYZlt{}Description}\PY{n+nt}{\PYZgt{}}Broker test event generated by Comet 1.1.0.\PY{n+nt}{\PYZlt{}/Description\PYZgt{}}
    \PY{n+nt}{\PYZlt{}Reference} \PY{n+na}{uri=}\PY{l+s}{\PYZdq{}http://comet.transientskp.org/\PYZdq{}}\PY{n+nt}{/\PYZgt{}}
  \PY{n+nt}{\PYZlt{}/What\PYZgt{}}
\PY{n+nt}{\PYZlt{}/voe:VOEvent\PYZgt{}}
\end{Verbatim}

\vspace{-15pt}
\hrulefill
\caption{An example of the form of VOEvent used for benchmark testing. The
\texttt{ivorn} attribute of the \texttt{VOEvent} element and the \texttt{Date}
element were automatically generated and reflect the time at which the packet
was created.}
\label{lst:testmessage}
\end{listing*}

The contents of a particular VOEvent document are not relevant when working
with the event database: the database operations only involve manipulating the
arrival time of the VOEvent and it's SHA-1 hash. For this test, therefore, we
do not make use of the corpus of events described above. Instead, a series of
test VOEvent packets of the form shown in Listing~\ref{lst:testmessage} was
generated. Each packet was compliant with the VOEvent 2.0 schema, but carried
a relatively small payload amounting to little more than a timestamp
reflecting when the event was created.

A batch of 10000 such test messages was generated and stored in memory. The
total time taken to both verify that each VOEvent was not initially present in
the event database and then record it in the event database was
recorded\footnote{In version 1.1.0 of Comet, as tested, checking and recording
an event are distinct operations. Later versions combine these to form an
atomic check-and-record operation, which is both improves performance and
avoids a race condition.}. Comet does not provide an interface to the event
database which does not involve calculating a SHA-1 hash; the time measured
therefore includes hash calculation for each event.

The experiment described was initially performed with the event database
stored on magnetic disk. The mean time taken to check and record an event in
the database is shown in Tab.~\ref{tab:perf:individual}. Note that this is
orders of magnitude above the times measured for the other processing steps.
This is, perhaps, unsurprising: accessing disk storage involves significant
overhead. To mitigate this, memory-based filesystem was created based on
\texttt{tmpfs} \citep{Kerrisk:2014} and both broker and subscriber were
configured to store their event databases here. This storage is entirely
RAM-based, so avoids the extra delays in writing to disk.

The experiment was repeated with the database stored on the \texttt{tmpfs}
filesystem; the result was a factor of 25 improvement in the time taken to
process each event, as shown in Tab.~\ref{tab:perf:individual}.

As per \S\ref{sec:design:dedup}, Comet stores hashes of the VOEvents received
for 30 days. On a busy VOEvent network, this could involve generating a much
larger database than the 10000 events tested, which may impact performance.
The previous experiment was therefore repeated 1000 times using the same
databases stored on \texttt{tmpfs}, resulting in a database containing $10^7$
hashes in total. The lowest mean processing time per event was measured when
processing batch 155, at an average of 0.000491\,s per event; the highest when
processing batch 855, at an average of 0.000502\,s per event. There was no
systematic increase in processing time with event database size. Testing with
a significantly larger database was impossible due to the available memory.

\subsubsection{XPath evaluation}
\label{sec:perf:individual:xpath}

The time taken to evaluate an XPath expression over a VOEvent document depends
not only on the complexity of the document being processed but also on the
XPath expression itself. A detailed discussion of the performance
characteristics of XPath is outside the scope of this work; instead, we take
the example queries given in \S\ref{sec:filter:xpath} as representative of a
typical workload.

All test VOEvent documents were read from disk, parsed, and stored in memory
using Comet's internal representation. Each of of the XPath expressions in
turn was parsed into an lxml \texttt{XPath} object, as used by Comet for XPath
filtering during normal operations. The total time taken to check all events
against each expression in turn was measured.

\subsubsection{Results}
\label{sec:perf:individual:results}

\begin{table}
\renewcommand*\footnoterule{}
\begin{minipage}{\columnwidth} 
\begin{center}

\caption{Timing results for each stage of Comet's processing of a VOEvent
document. All results except the check against the event database were based
on a corpus of 16425 genuine VOEvent documents; the check against the event
database was performed using synthetic test data. Each test is described in
\S\ref{sec:perf:individual}.}
\label{tab:perf:individual}
\begin{tabular}{lcc}
\toprule
\multicolumn{1}{c}{Operation} & \multicolumn{1}{c}{Total (s)} &\multicolumn{1}{c}{Per event (s)} \\
\midrule
XML parsing & 1.625465 & 0.000099 \\
SHA-1 calculation & 0.152024 & 0.000009 \\
\multicolumn{3}{l}{Event database operations:\footnote{Also includes SHA-1 calculation.}} \\
\hspace{4mm} Magnetic disk  & - & 0.013331 \\
\hspace{4mm} \texttt{tmpfs} & - & 0.000499 \\
Schema validation & 1.385420 & 0.000084 \\
\multicolumn{3}{l}{XPath evaluation:\footnote{Expressions as defined in \S\ref{sec:filter:xpath}.}} \\
\hspace{4mm}  Expression~\ref{xpath:exp1} & 0.218424 & 0.000013 \\
\hspace{4mm}  Expression~\ref{xpath:exp2} & 0.221899 & 0.000014 \\
\hspace{4mm}  Expression~\ref{xpath:exp3} & 0.579474 & 0.000035 \\
\hspace{4mm}  Expression~\ref{xpath:exp4} & 0.283376 & 0.000017 \\
\bottomrule
\end{tabular}
\end{center}
\end{minipage}
\end{table}

The total time for operating on all messages being tested (where applicable),
as well as the mean time per event, for each of the tests above is recorded in
Table~\ref{tab:perf:individual}.

Note that the results recorded for XPath filtering are not directly comparable
to those for the other tests described. All the other operations are performed
once per event received by the broker. In contrast, potentially several
different XPath expressions are evaluated per subscriber for every event
received. Thus, even though the time recorded for evaluating the XPath
expressions is substantially less than that recorded for event parsing or
schema validation, the total time spent on XPath processing may, in fact, be
greater in a deployed system.

Leaving aside XPath, of the individual operations performed once per event
interacting with the event database dominates: even when using a
\texttt{tmpfs}-backed database the time taken to check and record the event
hash is more than twice that spent on the other operations combined, and is
compounded by a further factor of over 25 when magnetic disks are used. Future
performance-focused development of Comet should investigate ways to mitigate
this issue.

\subsection{Latency}
\label{sec:perf:latency}

For certain science cases, maximizing the scientific relevance of follow-up
observations requires extremely rapid response. For example, identifying
precursors of fast radio bursts \citep{Thornton:2013} would require action on
a timescale a tens of milliseconds. It is therefore important that the VOEvent
transport system does not introduce excessive latency to the dissemination of
event notifications.

For the purposes of this discussion, we define the ``latency'' of a VOEvent as
the time elapsed between its creation by an author and the instant at which it
has been received by a subscriber and that subscriber is in a position to take
action (using the strategies described in \S\ref{sec:design:plugin}) based
upon it.

In this test, we measure the latency introduced by passing a VOEvent from an
author through a Comet broker and on to a Comet-based subscriber.

\subsubsection{Test setup}
\label{sec:perf:latency:setup}

A script was used to generate 3000 individual VOEvent packets of the form
shown in Listing~\ref{lst:testmessage} and submit them to a broker at
intervals of 0.3 seconds.

A plugin (\S\ref{sec:design:plugin}) was written which, whenever an event of
the form described above is received, compares the timestamp in the event with
the current time, and saves the difference to a log file. This plugin was
enabled on a subscriber, which was then connected to the broker.

Both the benchmarking script and the plugin described are available from the
Comet repository (\S\ref{sec:avail}).

\subsubsection{Results}
\label{sec:perf:latency:results}

\begin{figure}
  \begin{center}
  \includegraphics[width=\columnwidth]{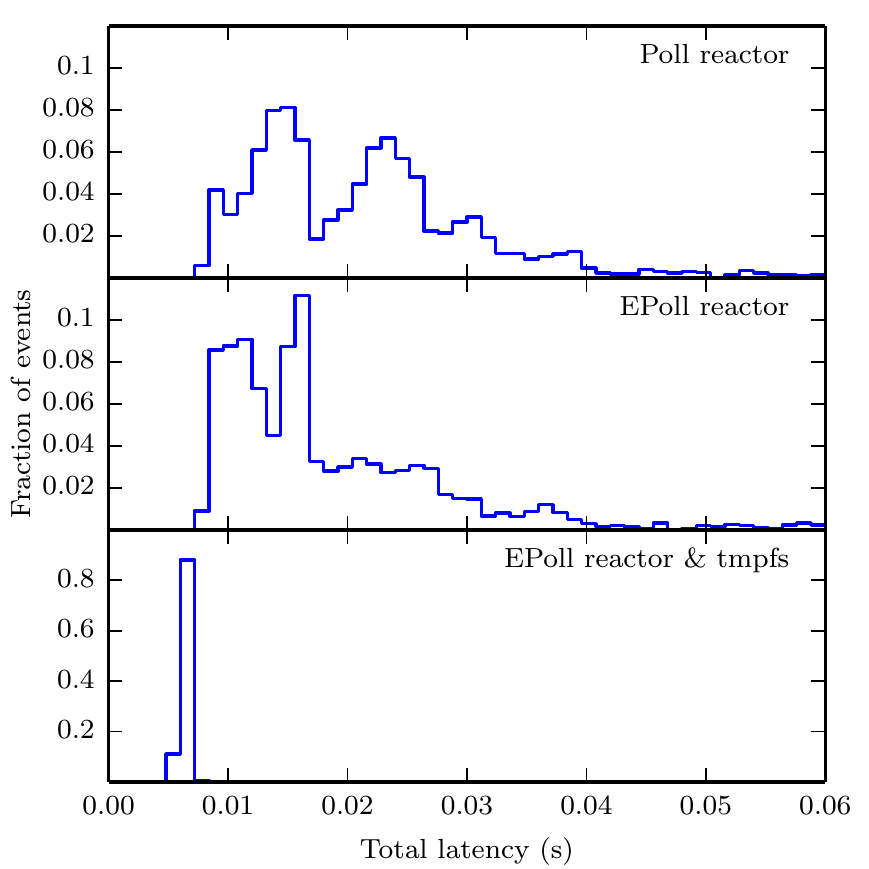}
  \end{center}

  \caption{Fraction of events received at a given latency (time between
  generation by the author and processing by the subscriber, as described in
  \S\ref{sec:perf:latency}). The uppermost plot reflects the default
  configuration; the central plot uses Twisted's \texttt{epoll()} based
  reactor; the bottom plot uses \texttt{epoll()} and stores the event database
  in memory.}

  \label{fig:latency}
\end{figure}

The distribution of latencies among the received events when this test was run
in the default configuration is shown in the top panel of
Fig.~\ref{fig:latency}. The mean latency was 0.022\,s with a standard
deviation of 0.011\,s; the longest recorded latency for any event was
0.171\,s.

As described in \S\ref{sec:design:twisted}, Twisted provides an event-driven
framework. The core of this framework is the ``reactor'', which provides a
uniform interface to event handling across the platforms upon which Twisted
can run. The internal implementation of the reactor itself can vary from
platform to platform to most efficiently take advantage of the facilities
available to it.

The default reactor implementation used by Twisted on the system used for
testing is based on the \texttt{poll()} system call \citep{Posix1:2013}.
However, modern Linux systems provide the alternative \texttt{epoll()} call
\citep{Kerrisk:2014} which provides a more efficient alternative. Twisted
provides a reactor which is based upon \texttt{epoll()}. The same experiment
was therefore repeated, but with both broker and subscriber based on this
alternative reactor. The results are shown in the central panel of
Fig.~\ref{fig:latency}. This provided a somewhat improved mean latency of
0.019\,s with a standard deviation of 0.011\,s, and a reduced maximum latency
of 0.130\,s.

Section~\ref{sec:perf:individual:eventdb} established that the event database
operations take an average of 0.013\,s when the database is stored on magnetic
disk, as it was in this default configuration: this is some 70\,\% of the
measured event latency. The same section demonstrated a 25-fold improvement
when the database was stored in RAM using the \texttt{tmpfs} filesystem. This
performance improvement comes at some cost: RAM technologies typically used in
modern systems are inherently volatile, and the event database would not
survive if the system were powered down or rebooted. Further, the $10^7$ event
database described in \S\ref{sec:perf:individual:eventdb} consumed around
1\,GiB of storage; given a database retention period of 30 days
(\S\ref{sec:design:dedup}) and potentially multi-million-per-day event rates
from next generation facilities, memory capacity may be a limiting factor.

These considerations notwithstanding, the database was re-created on a
\texttt{tmpfs} filesystem and the test repeated. The results are shown in the
bottom panel of Fig.~\ref{fig:latency}. Not only are these latencies lower
(mean 0.0063\,s, maximum 0.013\,s) than those based on magnetic disks, but
they are also much more consistent than the previous tests (standard deviation
0.00033\,s).

An overhead of no more than around ten milliseconds is comparable to that
which might be expected from network delays over short links, and is unlikely
to be of significance in all but the most demanding of astronomical
applications. Note, however, that this figure was measured on an otherwise
unloaded system: while it sets a lower bound on the latency added by Comet, a
production system under load is unlikely to perform at the same level.

For the rest of the tests presented in this manuscript, we continue to adopt
the \texttt{epoll()} and \texttt{tmpfs} configuration described here.

\subsection{Number of subscribers}

In order to meaningfully act as a distribution, rather than simply a
forwarding, system, and certainly in order to enable the construction of
extended networks of interconnected brokers, it is necessary that a single
Comet broker be able to serve many subscribers simultaneously. Here, we
measure how latency increases as more subscribers are connected to the the
broker.

\subsubsection{Test setup}

Using the same script as described in \S\ref{sec:perf:latency:setup}, 1000
test events were submitted to a broker. The number of clients connected to
that broker was increased at logarithmic intervals (1, 2, 4, ...). Each client
recorded the latency of each event received to a log file.

Each Comet process takes approximately 32\,MB of memory, used to hold the
Comet code itself, the associated libraries, the Python interpreter, and the
overhead associated with the Docker container. The test system contained 8\,GB
RAM\@. When testing with 256 subscribers, the machine ran out of memory and
started to swap to disk. This set an upper bound on the number of subscribers
which could be tested.

\subsubsection{Results}

The latency increases gradually with increasing subscriber count, from a mean
of 0.0063\,s for a single subscriber to 0.0931\,s for 256 subscribers, the
highest number tested: even at this level, the mean latency was less than
0.1\,s. The maximum latency rose to a peak of 0.49\,s.  A latency of around
0.1\,s is comparable to a long range (e.g.\ transatlantic) network round trip
times and is at a level where it may start to impact on time-critical
astronomical applications.

\begin{figure}
  \begin{center}
  \includegraphics[width=\columnwidth]{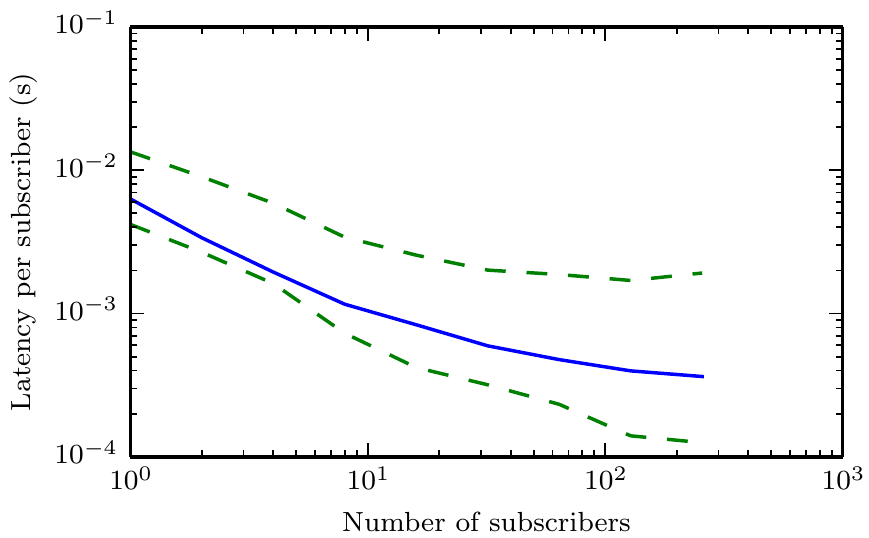}
  \end{center}

  \caption{Scaling of event latency as a function of subscriber count. The
  solid line shows the mean latency divided by the number of subscribers; the
  dashed lines are similar, but for the maximum and minimum latencies
  recorded.}

  \label{fig:subscribers}
\end{figure}

The scaling of latency with subscriber count is shown in
Fig.~\ref{fig:subscribers}. Note that the scaling is better than linear across
the range of subscriber counts tested: ingestion of new events into the
broker, rather than distribution to subscribers, dominates.

Other than a slowly increasing latency, the Comet system showed no
ill effects of handling a large number of subscribers: neither memory nor CPU
usage of the broker showed excessive growth. If the latency were acceptable
for the science application, there would be no difficulty in serving 256
subscribers in a production mode using this hardware.

For servicing extremely large numbers of clients while minimizing latency, a
tree-like structure could be established. For example, serving 8 subscribers
introduced a mean latency of 0.0093\,s. If each of those 8 subscribers
redistributed the event to a further 8 clients, we might expect a total
latency on the order of 0.02\,s to reach 256 clients; if the tree were
extended to ten levels we might expect to reach $8^{10}$ ($\sim10^9$)
subscribers with 0.1\,s latency.

\subsection{Total throughput}
\label{sec:perf:total}

Next-generation facilities will announce transients at rates far outstripping
those seen at present. Notably, LSST is predicted to reach an average rate of
$10^7$ events per night: assuming those events are evenly spaced over a 12
hour period, this is equivalent to over 230 events per second. This is the
output from just a single instrument, albeit a prolific one, and takes no
account of the cascade of follow-up packets that a significant transient would
likely generate. Here, we measure how the event rate processed by the Comet
broker running on the test system.

\subsubsection{Test setup}
\label{sec:perf:total:setup}

A script was used to generate 10000 individual VOEvent messages, which were
stored in RAM\@. After all of the events had been generated, the author
started submitting them to a Comet broker which had a single subscriber
attached. The total time taken by the author from the start of the submission
of the first event to the closing of the connection after the submission of
the last event was measured by recording its running time. The total time from
the receipt of the first event to the receipt of the last event by the
subscriber was measured by taking the difference between the latest and the
earliest timestamps recorded in the event database (\S\ref{sec:design:dedup}).
These times are then converted into an per-second event rate.

The number of concurrent connections between the author and the broker was
varied logarithmically. For each number of connections, the experiment was
repeated 10 times.

\subsubsection{Results}
\label{sec:perf:total:results}

\begin{figure}
  \begin{center}
  \includegraphics[width=\columnwidth]{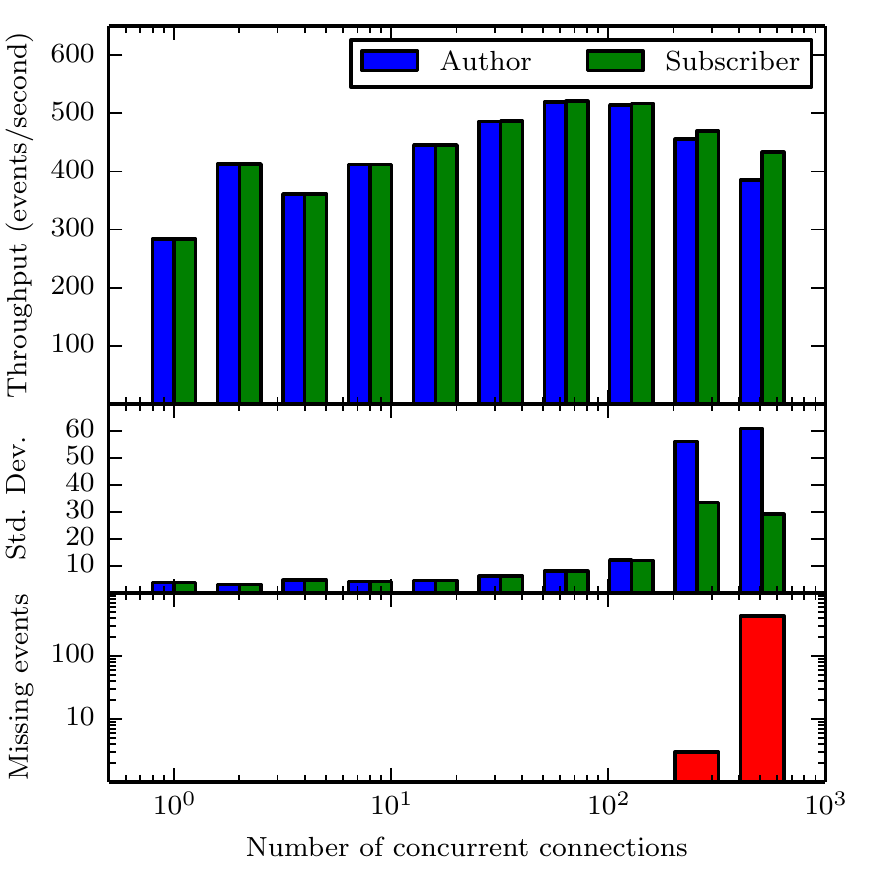}
  \end{center}

  \caption{At top, the mean throughput of events as transmitted by
  the author and as received by the subscriber as a function of number of
  concurrent connections from author to broker. The central panel shows the
  standard deviation of the measured rates. The number of events which were
  not successfully received by the subscriber is shown at the bottom.}

  \label{fig:throughput}
\end{figure}

Figure~\ref{fig:throughput} shows the how the event rate measured at both
author and subscriber varies with the number of concurrent connections. With a
single connection a rate of 283.7 events/second at the author and 283.8
events/second at the subscriber is achieved. This increases to a peak of 519.2
events/second at the author and 534.2 events/second at the subscriber with 64
concurrent connections; after this, increasing he number of connections causes
the overall throughput to drop. The standard deviation of the measured rate is
also plotted: the throughput is relatively stable at low connection counts,
but substantial variations are seen with 256 and 512 concurrent connections.

At the highest connection counts, the rate is not only seen to drop
substantially, but also some events are lost in transit: at the end of the
test, the subscriber had received fewer than 10000 VOEvent packets. Since the
throughput was lower at these rates, and since reliable transmission is
essential, concurrency levels higher than 512 connections were not
investigated.

These results may be explained by considering the balance between the
per-connection overhead and the compute load on the broker. Since each event
is delivered by making a new connection to the broker (as per the protocol
described in \S\ref{sec:vtp}) there is a per-event overhead due to creating
and tearing down the connection (\S\ref{sec:perf:highlatency} discusses some
of the overheads in managing TCP connections). At low connection counts, this
latency dominates; as the concurrency increases, the throughput is dominated
by the broker load.

At the highest connection counts, the configuration of the Linux kernel's
networking stack comes in to play. Large numbers of short lived connections
are a relatively uncommon phenomenon, and the standard configuration of the
Linux kernel is not optimized to handle them efficiently. Indeed, at very high
connection counts, the kernel logged warnings that it was was under a
``\textsc{syn} flood'' attack \citep{CERT:1996}. Under this load, connections
may be dropped or rejected by the kernel, leading to events never reaching
their destination, as seen in the lowest panel of Fig.~\ref{fig:throughput}.
Many options within the kernel may be tuned to improve its performance under
these network loads. However, since the peak throughput was already limited by
Comet's CPU requirements at lower connection counts, they were not
investigated here.

It is worth noting that, at low connection counts, the throughput from author
to broker and from broker to subscriber were effectively identical, but they
began to diverge as the concurrency increased. This is again due to the
per-connection overhead: since the connection from the broker to the
subscriber is permanently kept open, it is significantly more efficient, and
provides a continuously-available high bandwidth connection. At high
connection counts, the latency involved in servicing many connections means
that such high bandwidth cannot be achieved here when submitting events.

Without special tuning, a throughput of over 500 events per second is more
than twice that required to service the average event rate predicted from
LSST\@. Further, this test was limited by CPU performance on desktop-class
hardware that will be substantially more than a decade old before LSST is
commissioned. In these terms, then, servicing an LSST-scale event stream with
a VTP based broker seems plausible, although there are a number of caveats:

\begin{itemize}

  \item{This calculation takes no account of follow-up traffic generated in
  response to the events;}

  \item{These events did not carry a scientific payload, and hence are likely
  to be significantly smaller than those which might be transmitted in
  practice;}

  \item{Although the mean event rate from LSST will be around 250
  events/second, this will be transmitted in short bursts of much higher
  rates. Averaging the event traffic over time reduces the instantaneous
  traffic to a manageable level, but introduces significant additional
  latencies.}

\end{itemize}

\subsection{High-latency connections}
\label{sec:perf:highlatency}

Astronomical observatories are frequently located in remote locations: in
deserts, on mountain tops, and so on. The geographic isolation of these
facilities often results in their having poor internet connections. Even if
high-bandwidth networking is arranged specifically to service the observatory,
network latencies are likely to be high.

One might imagine that some preliminary data analysis for such an observatory
would be performed on-site, rather than attempting to ship large volumes of
raw data out of a remote location. Further, it would not be practical for
large numbers of external clients to connect inwards to a VTP broker running
at the observatory. Therefore, for the purposes of this discussion, we we
assume that the events are generated by a VOEvent author on site, then shipped
using VTP to a remote broker for public distribution.

Assuming 10 million alerts are issued by the observatory per night and each
event has a size of around 10\,kiB, a total of 100\,GiB of event data might be
created. Given that long range multi-gigabit per second connections are widely
available, the total amount of data to be transmitted is unlikely to be
intractable.

Network latency, however, presents a further problem. As described in
\S\ref{sec:vtp}, each event must be submitted by the author initiating a new
connection to the broker, submitting the event, waiting for an
acknowledgement, and then closing the connection. Sending the event and
waiting for acknowledgement involves a network round-trip.  However, data is
transmitted over TCP \citep{Cerf:1974}, we use the standard TCP mechanisms for
creating and terminating connections, each of which involves another network
round trip. This process is illustrated in Fig.~\ref{fig:tcp}: the complete
transaction involved in submitting a single event to the broker, given a
network round trip time of $t_\mathrm{RT}$, takes $3 t_\mathrm{RT}$. In
practice, after transmitting the final \textsc{fin} packet, the author may
assume that the connection is closed without waiting for a response, and hence
initiate a new connection, so the figure of $2 t_\mathrm{RT}$ describes the
interval between connection attempts.  Assuming that events are sent in
sequentially, and given a round trip time of, say, 500\,ms, this would limit
the rate at which events can be sent to 1 per second, or 43200 in a 12 hour
period. This is far short of the throughput discussed in
\S\ref{sec:perf:total}, and certainly inadequate for the putative 10 million
alerts per night discussed above. This is a significant flaw in the VTP
system. It is to be hoped that future revisions can address the issue; for
further discussion, see \S\ref{sec:future:bulk}.

Until and unless this problem is addressed, it is necessary to consider
alternative approaches. As discussed in \S\ref{sec:perf:total}, it is possible
for the author for an author to submit multiple events simultaneously by
opening more than one TCP connection to the broker. Here, we investigate to
what extent this can mitigate the issue.

\subsubsection{Test setup}

\begin{figure}
  \begin{center}
  \includegraphics[width=0.7\columnwidth]{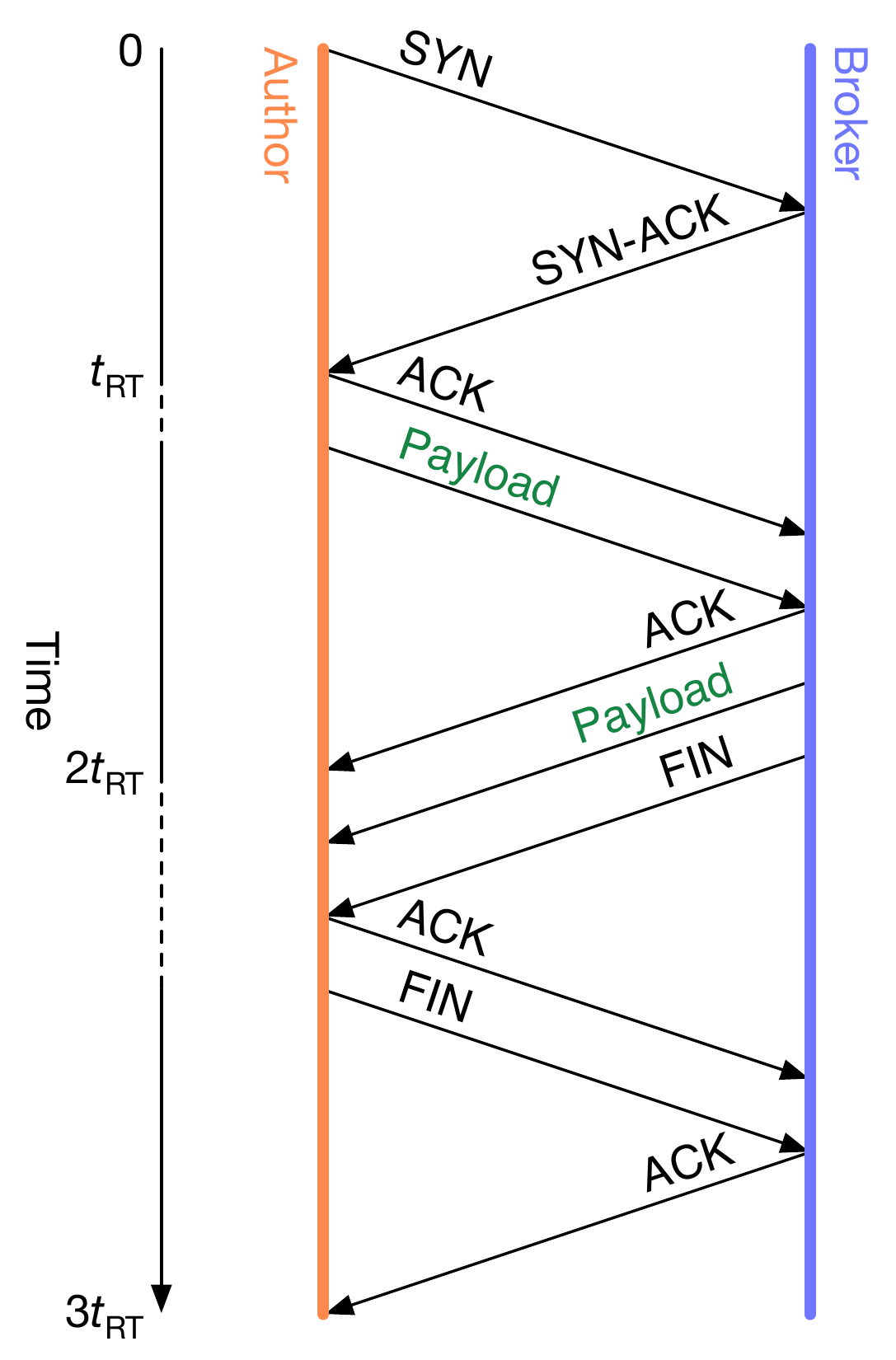}
  \end{center}

  \caption{The complete network packet exchange as an author uploads a VOEvent
  to a broker. Packets with specific TCP flags set have those flags indicated
  in upper case. Payload data (a VOEvent or Transport packet) are indicated
  by the word ``Payload''. Time increases down the diagram. The network round
  trip time is denoted by $t_\mathrm{RT}$. A dashed time axis indicates
  packets being sent with no interval between them. For example, at time
  $t_\mathrm{RT}$ the author sends \textsc{ack} immediately followed by
  VOEvent data.}

  \label{fig:tcp}
\end{figure}

Events were generated, sent the the broker, and thence onward to a single
subscriber as per \S\ref{sec:perf:total:setup}, and was carried out as
described in that section. Connection counts were again increased
logarithmically.  Given the relative stability of the throughput (at least for
modest connection counts) shown in Fig~\ref{fig:throughput}, a single set of
10000 events was sent for each level of concurrency.

Link-level network latency was simulated using NetEm \citep{Hemminger:2005},
the network emulation functionality available as part of the Linux kernel.
Given a (virtual, in this case) network device named \texttt{vethXXX}, a
delay of \texttt{YYY}\,ms may be added to each packet sent through it by
running:
Note that this delay applies only to packets sent through the interface: no
delay is applied to packets received by the interface. To simulate a symmetric
network delay using this approach, it would therefore be necessary to add a
latency of $t_\mathrm{RT} / 2$ at both the author and the subscriber
interfaces. However, this is complicated in the test system since the
subscriber also communicates with the broker over its interface. Therefore,
instead the whole delay was applied to the output of the author. Given the
symmetric nature of Fig.~\ref{fig:tcp}, the observed effect is identical.

\subsubsection{Results}

\begin{figure}
  \begin{center}
  \includegraphics[width=\columnwidth]{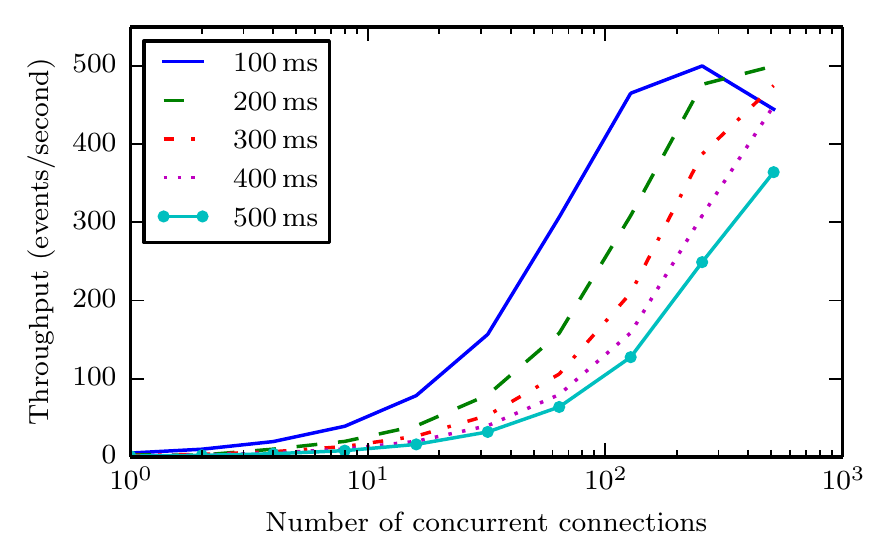}
  \end{center}

  \caption{Variation of throughput, as measured by the rate of events received
  by the subscriber, as a function of number of concurrent connections shown
  for a variety of network round trip times.}

  \label{fig:rtt}
\end{figure}

Figure~\ref{fig:rtt} shows how the throughput varies with the number of
concurrent connections for a variety of network round trip times. As expected,
at low concurrencies, the throughput is extremely low: the network round trip
time completely dominates the transmission rate. However, this is
substantially mitigated by increasing the concurrency: with a round trip time
of 100\,ms, using 256 concurrent connections provides a rate of 500
events/second, which approaches the peak rate achieved in in
\S\ref{sec:perf:total:results}. At higher concurrencies, though, the rate
begins to diminish as the load incurred in managing connections dominates,
as seen in \S\ref{sec:perf:total:results}.

A similar pattern is seen for other round trip delays: increasing the number
of connections can mitigate the effects of network-induced latency. However,
attempting to initiate more than 512 connections always resulted in a large
number of events getting dropped in transit, as the kernel refused to service
so many simultaneous network connections. Thus, the peak rates achieved at the
higher round-trip times were always suppressed relative to the throughput
measured with no latency. As discussed in \S\ref{sec:perf:total:results},
appropriate tuning of the kernel networking stack could be used to help
overcome this issue; however, a better approach would be to address it at the
protocol level, an idea to which we return in \S\ref{sec:future}.

\section{Authentication}
\label{sec:security}

For many applications involving VOEvents, it is important to be certain of the
authenticity of the event. That is, to be able to guarantee that the event
genuinely describes the results of observations by its supposed author. This
is important both for event authors, to protect their reputation for issuing
high quality, trustworthy events, and to subscribers, who cannot run the risk
of using expensive facilities chasing phantoms. While the overt motivation for
forging events is low---there is no obvious way to exploit a VOEvent for
monetary gain, for example---the potential for mischief-makers to play havoc
with event networks cannot be ignored.

Two approaches may be taken to securing an event distribution system. The
first is to authenticate the transport layer using a technology such as TLS
\citep{Dierks:2008}. In this way, each entity involved would be able to verify
both the integrity of a VTP connection and the identity of their remote peer.
A subscriber could therefore be certain of the identity of the broker from
which it receives a particular event. However, that broker was not itself the
originator of the event, but rather it received it either from the author
directly or from another broker: it is now incumbent upon that broker to not
only to verify the identity of the sender but also to satisfy the subscriber
that this has been done with sufficient diligence. If the event has traversed
a length path through multiple brokers before reaching the subscriber, this
task becomes prohibitively complex. As such, this is not a mechanism which VTP
supports.

The alternative is to authenticate individual VOEvent packets. This can be
done by applying a cryptographic signature to the event using a technology
such as OpenPGP \citep{Callas:2007} or XML Digital Signatures
\citep{Bartel:2008}. The recipient of an event can then verify that it is
identical to the event to which the signature was originally applied.

Work has already been carried out on applying XML Digital Signatures to
VOEvents \citep{Allen:2008} outside the framework of VTP\@. However, the
implementation is relatively complex: not only is there a paucity of libraries
providing a convenient implementation of the standard, but even the library
the authors chose to use\footnote{XMLSec;
\url{http://www.aleksey.com/xmlsec/}} required source-level modification to
meet their requirements.

On the other hand, both commercial and open-source implementations of OpenPGP
are widely available both as stand-alone tools and with programming language
interfaces. Furthermore, \citet{Denny:2008} describes a mechanism for
attaching an OpenPGP signature to a VOEvent with specific reference to VTP\@.
For these reasons, a prototype version of Comet with OpenPGP support has been
made available for testing.

\subsection{Implementation considerations}

The OpenPGP standard itself is widely used and tested: the basic cryptographic
guarantees it provides are as close to unimpeachable as it is reasonable to
ask for. However, there are three key hurdles which must be overcome before it
can be directly used in the context of VOEvents and VTP.

\subsubsection{Bitstream immutability}

Section~\ref{sec:design:dedup} discussed whether two VOEvent packets can be
regarded as ``the same'' and the motivated the requirement that entities
participating in a VTP network should transmit events unchanged. When
considering cryptographic signatures, this requirement becomes absolutely
fundamental. The signature is applied to a particular collection of bits, with
no semantic understanding of what those bits represent. If a single bit is
changed, the signature is invalidated, even if that change does not alter the
information content of the document and however inconsequential the change
might be.

Beyond its direct requirements on the transport layer, this could have
implications for various uses to which VOEvents may be put. For example, when
storing an event in an archival database, it would not be adequate to simply
extract the information from the packet and store that, re-serializing it to
XML if and when required. Rather, it would be necessary for the archive to
store the exact bitstream to which a signature had been applied.

\subsubsection{Event formatting}

The original proposal described by \citet{Denny:2008} makes use of the OpenPGP
cleartext signature framework. However, as \citet[\S7]{Callas:2007} makes clear, the
cleartext signature framework ``is not intended to be reversible'': in other
words, applying such a signature may modify the contents of the event packet
itself. Such modifications are generally insignificant (primarily concerning
the way in which lines starting with a ``-''---the ``hyphen-minus'' character,
Unicode code point U+002D---are handled), but, nevertheless, we regard any
mutation of the event data as unacceptable.

To avoid these proposals, we suggest adopting a modification of
\citeauthor{Denny:2008}'s proposal based on a detached signature
\citep[\S11.4]{Callas:2007} which is bundled with the VOEvent. It is this
modified proposal which is implemented in Comet.

\subsubsection{Trust model and key infrastructure}

Any entity can generate an OpenPGP key with whatever identifying name they
please and use it to apply a signature to a document. The recipient of the
document has a strong guarantee that the document was genuinely signed by the
given key, but has no particular reason to trust that the key was in the
possession of a reputable entity at the time the signature was made. At level,
subverting the system by signing VOEvents with valid-but-worthless keys
becomes a trivial exercise.

The most direct solution is for the owner of a key to directly provide it to
likely recipients in person or by some other tamper-proof means of
transmission. The recipient then knows that this particular key belongs to
that particular entity, and can take this into account when deciding whether a
signed event is genuine.

OpenPGP adopts extends this approach to the ``web of trust'' model. Here,
entities who have received a copy of the key directly from its owner can
themselves sign and redistribute it. The recipients can they choose whether
they believe the intermediary to be trustworthy to warrant the identity of the
owner. The recipients may sign and distribute the key further, eventually
building up a web of certified keys.

The same model may be applied to event packets themselves. Rather than simply
checking for a valid signature made by the author of the event, a legitimate
approach would be to check for a valid signature by any entity which the
recipient regards as trustworthy to guarantee the packet's authenticity. This
could include, for example, intermediate brokers or event aggregators.
However, this scheme is not provided for in the note by
\citeauthor{Denny:2008}, and has the significant downside of much increased
management overhead, particularly when automatic response to genuine events is
required: the recipient must indicate which entities they trust to sign events
from which authors.

\subsection{Usage in Comet}

The released version of Comet at the time of writing does not include support
for OpenPGP based event authentication. However, there is an experimental
version available which may be used for experimenting with these technologies.
See \S\ref{sec:avail} for information on how to obtain both released and
experimental versions of Comet.

Comet provides comprehensive support for all the modes in which event
authentication may be used within VTP\@. Specifically:

\begin{itemize}

  \item{When submitting to a broker, \textit{comet-sendvo} can apply a
  signature to the event being sent;}

  \item{When receiving an event from an author, the Comet can be set to
  only accept events which are appropriately signed;}

  \item{When receiving an event from a broker, Comet can be set to only act
  upon and redistribute events which are appropriately signed.}

\end{itemize}

Comet also supports subscriber authentication by applying the same signing
mechanisms to Transport documents (\S\ref{sec:vtp}). Using this technique:

\begin{itemize}

  \item{On receiving a connection from a subscriber, Comet can request that
  the subscriber authenticate themselves by means of a signed Transport
  message, and will then only distribute events to subscribers which provide
  trustworthy signatures.}

  \item{When subscribing to a remote broker, Comet can provide a signed
  Transport message in response to an authentication request.}

\end{itemize}

Comet's OpenPGP support is based upon GnuPG\footnote{\url{http://gnupg.org/}}.
Comet does not provide any mechanism for managing the configuration of GnuPG:
instead, the standard GnuPG tools should be used for this, and Comet inherits
the configuration and key database from them.

Of course, generating and verifying a cryptographic signature requires some
numerical calculation. Furthermore, for security reasons, directly linking
GnuPG as a library in application code is not supported. Handling
cryptographic operations in-process is therefore not possible. Instead, it is
necessary to fork a separate GnuPG process, incurring additional overhead.
Therefore, the impact of OpenPGP support on Comet's performance must be
considered.

In practice, the overhead of signing an event is insignificant: any one author
is likely to be generating only a limited number of events, and, even if that
number is large, they can trivially spread the load across multiple machines.
However, the Comet broker must check the signatures of all events received: it
is here that performance issues become critical.

A simple test was performed to measure the time taken to check the signature
on a VOEvent packet. 1000 distinct VOEvent packets of the form shown in
Listing~\ref{lst:testmessage} were generated and signed using the Comet
codebase. Each signature in turn was then checked for validity. The total
time taken to check all signatures on the system described in
\S\ref{sec:perf:system} was 22.90\,s, or around 0.023\,s per event. This is
broadly comparable to values which might be expected due to network latency,
and is a factor of $\sim3.6$ greater than the latency introduced by the Comet
broker when not checking a signature (\S\ref{sec:perf:latency:results}). While
not prohibitively expensive, then, the overhead introduced by this technique
cannot be ignored by administrators of heavily-loaded brokers.

\section{Future VOEvent and VTP revision}
\label{sec:future}

This manuscript has described both VTP itself and the issues that have arisen
when developing a specific implementation of it.  From these considerations,
five specific recommendations for future evolution of the VOEvent and VTP
standards can be drawn. Some of these will be incorporated into a revised
version of VTP which will be submitted for IVOA standardization at a later
date.

\subsection{Event identity}
\label{sec:future:identity}

Section~\ref{sec:design:dedup} discussed the question of the identity of a
VOEvent. In particular, it considered whether two events encoding identical
information but in with a different (perhaps only marginally) serialization
could be regarded as the same event. This is not well defined by the current
VOEvent standard \citep{Seaman:2011}.

As discussed, the question of the identity of events is important to the
implementation of VTP networks. However, it is also of wider relevance: the
VOEvent identifier provides a convenient means to refer to a particular
celestial transient in a variety of context, but can only be reliably used as
such if it is unambiguously defined.

\subsection{Packet immutability}
\label{sec:future:immutability}

It is an implicit requirement of VTP and of event authentication techniques
based on OpenPGP signatures that the \textit{bitstream} of a packet must be
unchanged by the process of transmission over VTP\@. This requirement goes
beyond the straightforward requirement that the \textit{information contained
within} an event must be unchanged. The more stringent requirements of VTP are
not explicit in the current version of the protocol definition.

\subsection{Event de-duplication}
\label{sec:future:dedup}

Section~\ref{sec:design:dedup} described de-duplication to avoid loops on a
VTP network. This requirement is not explicit within the current VTP
definition. Comet has demonstrated an effective approach to this problem
building upon \S\S\ref{sec:future:identity} and \ref{sec:future:immutability}.

\subsection{Filtering}
\label{sec:future:filter}

Section~\ref{sec:filter} demonstrated that the design of VTP is easily
extensible to accommodate relatively complex broker-side filtering
capabilities. However, the implementation of these filters in Comet requires a
non-standard extension to the protocol. Future VTP revisions should consider a
formalized means of enabling brokers to advertise what filtering capabilities
they are capable of providing, if any, and for subscribers to specify any
filters required.

\subsection{Bulk event submission}
\label{sec:future:bulk}

Section~\ref{sec:perf} demonstrated that Comet was capable of receiving and
distributing large numbers of events with relatively low latency. However,
\S\S\ref{sec:perf:total} and \ref{sec:perf:highlatency} demonstrated that
the major limiting factor on performance, in particular in the case of high
network round trip times, is the requirement that each individual event
submission by an author take place over a new TCP connection.

Two approaches should be considered to this flaw in the protocol. The first is
simply to drop the requirement that the connection should be closed between
each submission. Not only would this reduce the total transaction time per
event by removing the need to repeat the TCP handshake (see
Fig.~\ref{fig:tcp}), it would also be possible to interleave transactions: the
author could begin the submission of further events before having received an
acknowledgement of the first.

The alternative approach is to group batches of events into a single data
structure (a ``container''), and transmit that over VTP in a single
transaction. The definition of a container format for VOEvents is already
under discussion in the context of the
IVOA\footnote{\url{http://www.ivoa.net/forum/voevent/2013-November/002914.html}}.

\section{Availability}
\label{sec:avail}

Comet is freely available, open source software released under a two-clause
BSD-style\footnote{\url{http://opensource.org/licenses/BSD-2-Clause}} license.
It includes a comprehensive test suite and documentation. It is developed
using a public code repository; contributions and bug reports are actively
solicited.  Further details, including download and installation instructions,
are available from the project
website\footnote{\url{http://comet.transientskp.org/}}.

All materials used to generate this manuscript, including the Docker
configuration, benchmarking scripts, and latency measurement plugin are
available from the Comet repository.

\section{Conclusions}
\label{sec:conclusions}

The VOEvent Transport Protocol is an intentionally minimal mechanism for
distributing notifications of transient celestial events in the form of
VOEvent messages. Comet has been developed to implement all the core aspects
of VTP\@. It is production-ready software, and is freely available and ready to
be integrated into a variety of scientific projects.

This manuscript has described how Comet has been designed to meet the
requirements of VTP based upon an asynchronous, event-driven style of
programming. This has made it possible to provide a robust, high-performance
and easily extensible implementation of the protocol. The development of Comet
cast light on a number of areas of the protocol and of the wider VOEvent
infrastructure where additional clarity and specification is required.

Using Comet as a test-bed, we have investigated the performance
characteristics of VTP under a variety of conditions. Our results demonstrate
that VTP is broadly capable of meeting the anticipated requirements of the
next generation of large scale transient survey projects. However, there are
deficiencies in the design of the protocol which adversely affect its
perfomance. We have discussed how future revisions of VTP could address these
problems. We have also shown a prototype of a highly-configurable event
filtering system which will enable end users to sift through high-volume event
streams and receive only those events which are of relevance to their own
scientific goals.

\section{Acknowledgements}
\label{sec:ack}

The author is grateful to Bob Denny and Alasdair Allan for many useful
discussions on the design and implementation of the VOEvent Transport Protocol
and to Tim Staley and Roy Williams for their feedback on the design and
capabilities of Comet.

Both Comet itself and the tests described herein rely on a number of open
source software packages.  Python\footnote{\url{https://www.python.org/}}
provides both a convenient development platform and a rich variety of
libraries upon which to build; it is these, notably Twisted, lxml,
zope.interface\footnote{\url{http://docs.zope.org/zope.interface/}} and
ipaddr-py\footnote{\url{https://code.google.com/p/ipaddr-py/}} which have made
the development of Comet possible. Additionally, the support for event
authentication described \S\ref{sec:security} makes use of GnuPG and
PyGPGME\footnote{\url{https://launchpad.net/pygpgme}}. Docker is an invaluable
aid to testing, benchmarking and deploying Comet.

The author acknowledges support from the European Research Council via
Advanced Investigator Grant 247295.

\section*{References}

\bibliographystyle{elsarticle-harv}
\bibliography{comet}

\begin{thebibliography}{28}
\expandafter\ifx\csname natexlab\endcsname\relax\def\natexlab#1{#1}\fi
\expandafter\ifx\csname url\endcsname\relax
  \def\url#1{\texttt{#1}}\fi
\expandafter\ifx\csname urlprefix\endcsname\relax\def\urlprefix{URL }\fi

\bibitem[{{Allan} and {Denny}(2009)}]{Allan:2009}
{Allan}, A., {Denny}, R.~B., Aug. 2009. {VOEvent Transport Protocol, Version
  1.1}. {Note}, {International Virtual Observatory Alliance}.

\bibitem[{{Allen}(2008)}]{Allen:2008}
{Allen}, S.~L., Mar. 2008. {VOEvent authentication via XML digital signature}.
  Astron. Nachr. 329, 298--300.

\bibitem[{{AT\&T}(1979)}]{DBM:1979}
{AT\&T}, 1979. {Unix Programmer's Manual, Seventh Edition, Volume 1}.

\bibitem[{Bartel et~al.(2008)Bartel, Boyer, Fox, LaMacchia, and
  Simon}]{Bartel:2008}
Bartel, M., Boyer, J., Fox, B., LaMacchia, B., Simon, E., 2008. {XML Signature
  Syntax and Processing (Second Edition)}. {Recommendation}, {World Wide Web
  Consortium}.

\bibitem[{Bray et~al.(2008)}]{Bray:2008}
Bray, T., et~al., 2008. {Extensible Markup Language (XML) 1.0 (Fifth Edition)}.
  {Recommendation}, {World Wide Web Consortium}.

\bibitem[{Callas et~al.(2007)Callas, Donnerhacke, Finney, Shaw, and
  Thayer}]{Callas:2007}
Callas, J., Donnerhacke, L., Finney, H., Shaw, D., Thayer, R., 2007. {OpenPGP
  Message Format}. {Request for Comments} 4880, {Internet Engineering Task
  Force}.

\bibitem[{Cert and Kahn(1974)}]{Cerf:1974}
Cert, V.~G., Kahn, R.~E., 1974. {A protocol for packet network
  intercommuncation}. {IEEE Transactions on Communications} 22, 637--648.

\bibitem[{Clark and DeRose(1999)}]{Clark:1999}
Clark, J., DeRose, S., 1999. {XML Path Language (XPath), Version 1.0}.
  {Recommendation}, {World Wide Web Consortium}.

\bibitem[{{Computer Emergency Response Team}(1996)}]{CERT:1996}
{Computer Emergency Response Team}, 1996. {TCP SYN Flooding and IP Spoofing
  Attacks}. CA-1996-21.

\bibitem[{Denny(2008)}]{Denny:2008}
Denny, R.~B., May 2008. {A proposal for digital signatures in VOEvent
  messages}. {Note}, {International Virtual Observatory Alliance}.

\bibitem[{Dierks and Rescorla(2008)}]{Dierks:2008}
Dierks, T., Rescorla, E., 2008. {The Transport Layer Security (TLS) Protocol,
  Version 1.2}. {Request for Comments} 5246, {Internet Engineering Task Force}.

\bibitem[{Eastlake and Jones(2001)}]{Eastlake:2001}
Eastlake, 3rd, D., Jones, P., 2001. {US Secure Hash Algorithm 1 (SHA1)}.
  {Request for Comments} 3174, {Internet Engineering Task Force}.

\bibitem[{Fuller et~al.(1993)Fuller, Li, Yu, and Varadhan}]{Fuller:1993}
Fuller, V., Li, T., Yu, J., Varadhan, K., 1993. {Classless Inter-Domain Routing
  (CIDR): an Address Assignment and Aggregation Strategy}. {Request for
  Comments} 1519, {Internet Engineering Task Force}.

\bibitem[{Gau et~al.(2012)}]{Gau:2012}
Gau, S., et~al., 2012. {W3C XML Schema Definition Language 1.1 Part 1:
  Structures}. {Recommendation}, {World Wide Web Consortium}.

\bibitem[{Hemminger(2005)}]{Hemminger:2005}
Hemminger, S., 2005. {Network Emulation with NetEm}. In: {Proceedings of
  linux.conf.au 2005}.

\bibitem[{{IEEE} and {The Open Group}(2013)}]{Posix1:2013}
{IEEE}, {The Open Group}, 2013. {POSIX.1-2008}.

\bibitem[{Kantor(2014)}]{Kantor:2014}
Kantor, J., 2014. {Transient alerts in LSST}. In: {Proceedings of Hotwiring the
  Transient Universe 3}.

\bibitem[{Kerrisk et~al.(2014)}]{Kerrisk:2014}
Kerrisk, M., et~al., 2014. {Linux Programmer's Manual}. {The Linux man-pages
  project}, version 3.61.

\bibitem[{Lee(2006)}]{Lee:2006}
Lee, E.~A., May 2006. The problem with threads. IEEE Computer 39, 33--42.

\bibitem[{Peterson et~al.(2012)}]{Peterson:2012}
Peterson, D., et~al., 2012. {W3C XML Schema Definition Language 1.1 Part 2:
  Datatypes}. {Recommendation}, {World Wide Web Consortium}.

\bibitem[{Rixon(2005)}]{Rixon:2005}
Rixon, G., Aug. 2005. {Single-Sign-On Authentication for the IVO: introduction
  and description of principles, Version 1.00}. {Recommendation},
  {International Virtual Observatory Alliance}.

\bibitem[{{Seaman} et~al.(2006)}]{Seaman:2006}
{Seaman}, R., et~al., Nov. 2006. Sky event reporting metadata, version 1.11.
  {Recommendation}, {International Virtual Observatory Alliance}.

\bibitem[{{Seaman} et~al.(2011)}]{Seaman:2011}
{Seaman}, R., et~al., Jul. 2011. Sky event reporting metadata, version 2.0.
  {Recommendation}, {International Virtual Observatory Alliance}.

\bibitem[{Staley et~al.(2013)Staley, Titterington, Fender, Swinbank, van~der
  Horst, Rowlinson, Scaife, Grainge, and Pooley}]{Staley:2013}
Staley, T.~D., Titterington, D.~J., Fender, R.~P., Swinbank, J.~D., van~der
  Horst, A.~J., Rowlinson, A., Scaife, A.~M.~M., Grainge, K.~J.~B., Pooley,
  G.~G., 2013. {Automated rapid follow-up of Swift gamma-ray burst alerts at
  15\,GHz with the AMI Large Array}. Mon. Not. R. Astron. Soc. 428, 3114--3120.

\bibitem[{Swinbank(2014)}]{Swinbank:2014}
Swinbank, J.~D., 2014. {VOEvent: Were we are; where we're going}. In:
  {Proceedings of Hotwiring the Transient Universe 3}.

\bibitem[{{Thornton} et~al.(2013)}]{Thornton:2013}
{Thornton}, D., et~al., Jul. 2013. {A Population of Fast Radio Bursts at
  Cosmological Distances}. Science 341, 53--56.

\bibitem[{{Williams} et~al.(2012){Williams}, {Barthelmy}, {Denny}, {Graham},
  and {Swinbank}}]{Williams:2012}
{Williams}, R.~D., {Barthelmy}, S.~D., {Denny}, R.~B., {Graham}, M.~J.,
  {Swinbank}, J., Sep. 2012. {Responding to the Event Deluge}. Vol. 8448 of
  SPIE Conf. Ser.

\bibitem[{{Williams} et~al.(2009){Williams}, {Djorgovski}, {Drake}, {Graham},
  and {Mahabal}}]{Williams:2009}
{Williams}, R.~D., {Djorgovski}, S.~G., {Drake}, A.~J., {Graham}, M.~J.,
  {Mahabal}, A., Sep. 2009. {Skyalert: Real-time Astronomy for You and Your
  Robots}. In: {Bohlender}, D.~A., {Durand}, D., {Dowler}, P. (Eds.),
  Astronomical Data Analysis Software and Systems XVIII. Vol. 411 of
  Astronomical Society of the Pacific Conference Series. pp. 115--119.

\end{thebibliography}

\end{document}